\documentclass[twocolumn,journal]{IEEEtran}
\usepackage[T1]{fontenc}
\usepackage[latin9]{inputenc}
\usepackage{array}
\usepackage{verbatim}
\usepackage{prettyref}
\usepackage{multirow}
\usepackage{amsmath}
\usepackage{graphicx}
\usepackage[unicode=true,
 bookmarks=true,bookmarksnumbered=true,bookmarksopen=true,bookmarksopenlevel=1,
 breaklinks=false,pdfborder={0 0 0},pdfborderstyle={},backref=false,colorlinks=false]
 {hyperref}
\hypersetup{pdftitle={Your Title},
 pdfauthor={Your Name},
 pdfpagelayout=OneColumn, pdfnewwindow=true, pdfstartview=XYZ, plainpages=false}

\makeatletter

\providecommand{\tabularnewline}{\\}

\let\oldforeign@language\foreign@language
\DeclareRobustCommand{\foreign@language}[1]{%
  \lowercase{\oldforeign@language{#1}}}

\usepackage[caption=false,font=footnotesize]{subfig}
\usepackage{cite}
\usepackage{morefloats}
\usepackage{dblfloatfix}  

\@ifundefined{showcaptionsetup}{}{%
 \PassOptionsToPackage{caption=false}{subfig}}
\usepackage{subfig}
\makeatother

\begin{document}
\title{Utilizing Players' Playtime Records for Churn Prediction: Mining Playtime
Regularity}
\author{Wanshan~Yang,~\IEEEmembership{Student Member,~IEEE,} Ting~Huang,
Junlin~Zeng, Lijun~Chen,~\IEEEmembership{Member,~IEEE,} Shivakant~Mishra,~\IEEEmembership{Member,~IEEE}
and~Youjian~(Eugene)~Liu,~\IEEEmembership{Member,~IEEE}\thanks{Wanshan~Yang is with the Department of Computer Science, University
of Colorado Boulder, Boulder, USA, e-mail: \protect\href{mailto:wanshan.yang\%40colorado.edu}{wanshan.yang@colorado.edu}.}\thanks{Ting~Huang is with the Department of Data Analytics, Yoozoo Games,
Shanghai, China, e-mail: \protect\href{mailto:thuang\%40yoozoo.com}{thuang@yoozoo.com}.}\thanks{Junlin~Zeng is with the Department of Data Analytics, Yoozoo Games,
Shanghai, China, e-mail: \protect\href{mailto:zengjl\%40yoozoo.com}{zengjl@yoozoo.com}.}\thanks{Lijun~Chen is with the Department of Computer Science, University
of Colorado Boulder, Boulder, USA, e-mail: \protect\href{mailto:lijun.chen\%40colorado.edu}{lijun.chen@colorado.edu}.}\thanks{Shivakant~Mishra is with the Department of Computer Science, University
of Colorado Boulder, Boulder, USA, e-mail: \protect\href{mailto:mishras\%40colorado.edu}{mishras@colorado.edu}.}\thanks{Youjian~(Eugene)~Liu is with the Department of Electrical, Computer,
\& Energy Engineering, University of Colorado Boulder, Boulder, USA,
e-mail: \protect\href{mailto:eugeneliu\%40ieee.org}{eugeneliu@ieee.org}.}}
\IEEEaftertitletext{}
\markboth{}{}
\IEEEpubid{}
\maketitle
\begin{abstract}
In the free online game industry, churn prediction is an important
research topic. Reducing the churn rate of a game significantly helps
with the success of the game. Churn prediction helps a game operator
identify possible churning players and keep them engaged in the game
via appropriate operational strategies, marketing strategies, and/or
incentives. Playtime related features are some of the widely used
universal features for most churn prediction models. In this paper,
we consider developing new universal features for churn predictions
for long-term players based on players' playtime. %
In particular, we measure playtime regularity using the notion of
\textit{entropy} and\textit{ cross-entropy} from information theory.
After we calculate the playtime regularity of players from data sets of
six free online games of different types.%
{} We leverage information from players' playtime regularity in the
form of universal features for churn prediction. Experiments show
that our developed features are better at predicting churners compared
to baseline features. %
Thus, the experiment results imply that our proposed features could
utilize the information extracted from players' playtime more effectively
than related baseline playtime features.
\end{abstract}


\IEEEpeerreviewmaketitle{}

\section{Introduction\label{sec:Introduction}}

\IEEEPARstart{F}{ree} online games allow players to access games for free. As in other freemium
products and services, the revenue of a free online game company depends
on in-game purchases, and a larger player base indicates greater potential
revenue. Retaining current players is usually much easier and less
costly than recruiting new players. Therefore, these game companies
strive to identify potential churners in order to retain them via
proper operational strategies and incentive mechanisms.

Recent efforts on churn prediction in the game industry have employed
various methods and models, such as binary predictions, survival ensembles,
and the Cox model, and have utilized different features, including
playtime, login frequency, player in-game state, and player in-game
activity, such as purchases; see, e.g., \cite{Cowling_20152ICCIGC_PredictingPlayerDisengagementFirstPurchaseEventfrequencyBasedDataRepresentation,Bauckhage_20142ICCIG_PredictingPlayerChurnWild,Runge_2016ACS_RapidPredictionPlayerRetentionFreetoPlayMobileGames,Faltings_20142ICCIG_ChurnPredictionHighvaluePlayersCasualSocialGames,Bauckhage_20162ICCIG_PredictingRetentionSandboxGamesTensorFactorizationbasedRepresentationLearning,Hitchens_20162ICCIG_PredictingPlayerChurnDestinyHiddenMarkovModelsApproachPredictingPlayerDepartureMajorOnlineGame,Perianez_20172ICCIG_GamesBigDataScalableMultidimensionalChurnPredictionModel,Kim_20172ICCIG_ExtractingGamersCognitivePsychologicalFeaturesImprovingPerformanceChurnPredictionMobileGames,Kim_2018ITG_ProfitOptimizingChurnPredictionLongtermLoyalCustomerOnlineGames,Kim_2019ITG_GameDataMiningCompetitionChurnPredictionSurvivalAnalysisUsingCommercialGameLogData,Andjelkovic_2017ESwA_EarlyChurnPredictionPersonalizedTargetingMobileSocialGames-1,Perianez_2019ACS_UnderstandingPlayerEngagementInGamePurchasingBehaviorEnsembleLearning}.
In \cite{Wang_2018ACS_SemiSupervisedInductiveEmbeddingModelChurnPredictionLargeScaleMobileGames},
a churn prediction model is built based on user-app relationships
in a game launcher platform.

Playtime is widely selected as a universal churn prediction feature.
Unlike game-specific features that are different for different games
or may exist in one game but not in others, which limits their applicability,
playtime is one of the most fundamental records in any game database,
irrespective of game-specific characteristics. It is also more reliable
than other records, such as the players' login data that sometimes
depends on network conditions instead of the players' behavior. By
playtime, we mean the records of when a player play the game and for
how long.

However, there is a lack of research in further utilizing players'
playtime. Thus, in our previous work \cite{Liu_20192ICGC_MiningPlayerIngameTimeSpendingRegularityChurnPredictionFreeOnlineGames},
we considered utilizing the players' playtime to measure their playtime
regularity for churn prediction. Specifically, we considered long-term
players, who stay in the game for a sufficiently long period of time
and calculate the empirical distributions and related entropies of
these players' playtime. After observing the differences of related
entropies of these players' playtime between churners and non-churners,
we developed playtime related %
features for churner prediction based on these distributions and entropies.
However, when \cite{Liu_20192ICGC_MiningPlayerIngameTimeSpendingRegularityChurnPredictionFreeOnlineGames}
examined how each player compares to the entire game community that
he/she is playing with, the results varied if the proportion of churners/non-churners
changed in the dataset. Our previous work only tested the developed
playtime related features on two games in the same type and the generality
of our result might be a limit.

In this paper, we have improved the way of examining how each player
compares to the entire game community that he/she is playing with.
And we have tested our developed playtime related features for churner
prediction on more games of different types to show the generality.
We present a deeper and more thorough analysis of how churners and
non-churners differ in their playtime regularity, thereby expanding
our prior work \cite{Liu_20192ICGC_MiningPlayerIngameTimeSpendingRegularityChurnPredictionFreeOnlineGames}.

The main contributions of our work are listed below.%

\begin{itemize}
\item We have improved the model of long-term players' playtime regularity
based on the data of players' playtime distribution from \cite{Liu_20192ICGC_MiningPlayerIngameTimeSpendingRegularityChurnPredictionFreeOnlineGames}.
\item We inspect churners' and non-churners' evolvement of playtime regularity
across six free online games of different types. Then we propose features
based on players' playtime regularity for long-term players.%
\item We conduct experiments to evaluate our developed features across those
six free online games' data sets and show that these features could
help achieve a better prediction performance than the baseline features.
The experiments results imply that our proposed features could utilize
the information extracted from players' playtime more effectively
than the baseline playtime related features. \\
\end{itemize}
The rest of the paper is organized as follows. \prettyref{sec:Free-Online-Game}
describes the game data sets we use. \prettyref{sec:Representing-Player-Time}
explains how we split up the playtime of a player into periods and
how different distributions of players' playtime can be defined to
capture the players' playtime regularity of a player. \prettyref{sec:Mining-Player-Time}
illustrates how churners and non-churners evolve differently over
time. \prettyref{sec:Feature-Engineering} presents the process of
feature engineering from players' playtime regularity and \prettyref{sec:Churner-Detection-(Building}
evaluates the performance of our proposed features. \prettyref{sec:Conclusions}
concludes the paper.

\section{Free Online Game Data Sets\label{sec:Free-Online-Game}}

This work utilizes non-game-specific playtime data. The effectiveness
of the proposed methods are evaluated using the datasets of six free
online games: \textit{Thirty-six Stratagems (TS)}, \textit{Thirty-six
Stratagems} \textit{Mobile} \textit{(TSM), Game of Thrones Winter
is Coming (GOT), Womanland in Journey to the West (WJW), League of
Angels II (LOA II)},\textit{ }and\textit{ Era of Angels (EOA)}.

\subsection{Background of the Selected Games}

The above games are free online games published by Yoozoo Games.\footnote{https://www.yoozoo.com/}
These games are published on PC and mobile platforms and are designed
for different game types. The reasonable number of active players
and the diverse in-game mechanisms make the data from these six games
highly suitable to extract and evaluate our proposed features and
churn prediction algorithms.

\subsection{Data Selection}

Since a short-term players' playtime provides little information,
we consider those long-term players who have played the game for at
least $15$ days.

To define churners and non-churners in the free online games, notice
that the churners are unlikely to withdraw their accounts even if
they stop playing the game for a long time. We thus define a churner
as a player who does not access the game consecutively for a certain
number of days. Motivated by \cite{Kim_2018ITG_ProfitOptimizingChurnPredictionLongtermLoyalCustomerOnlineGames},
in these games, we define a churner as a player who stops playing
for more than 3 days, because we find that in our data sets, over
95\% of such players do not return to the game.

For our numeric analysis, we randomly select a set of long-term players
with the same numbers of churners and non-churners for each game.
The game platform, the game type, and the number of players, including
churners and non-churners for each game, are listed in Table. \ref{tab:Datasets-1}.\footnote{RPG represents role-playing game. MMORPG represents massively multiplayer
online role-playing game.}

\begin{table}[t]
\centering{}\caption{\label{tab:Datasets-1}Details of Selected Game Data Sets}
\begin{tabular}{|>{\centering}m{0.28\columnwidth}|>{\centering}m{0.1\columnwidth}|>{\centering}m{0.13\columnwidth}|>{\centering}m{0.12\columnwidth}|>{\centering}m{0.11\columnwidth}|}
\hline 
\centering{}Selected Games & Platform & Type & \#Churners & \#Non-churners\tabularnewline
\hline 
\textit{Thirty-six Stratagems (TS)} & PC & Strategy & 3596 & 3596\tabularnewline
\hline 
\textit{Game of Thrones Winter is Coming (GOT)} & PC & Strategy & 1716 & 1716\tabularnewline
\hline 
\textit{Thirty-six Stratagems} \textit{Mobile (TSM)} & Mobile & Strategy & 3062 & 3062\tabularnewline
\hline 
\textit{Womanland in Journey to the West (WJW)} & Mobile & RPG & 1702 & 1702\tabularnewline
\hline 
\textit{Era of Angels (EOA)} & Mobile & MMORPG & 1662 & 1662\tabularnewline
\hline 
\textit{League of Angels II (LOA II)} & Mobile & Card Game & 1872 & 1872\tabularnewline
\hline 
\end{tabular}
\end{table}

\section{Player Time Spending Distribution\label{sec:Representing-Player-Time}}

The players' playtime distribution describes how a player allocates
his/her time spent in a given game. In this section, we consider the
playtime distributions at different aggregation levels during the
latest playing periods of a player. As will be seen later, these distributions
will be the basis for the proposed feature engineering and churn prediction
method.

\subsection{Latest Playtime and Periods}

We consider the latest $n$ days of playtime of a player. Let $t_{u,d,r}$
be the playing time spent by player $u$, on day $d$, within hour
$r$, where $d=1,2,...,n$, and $r=1,2,...,24$. For example, for
$n=15$, if player 2 kept playing the game for 15 days continuously
and his/her latest playing date is December 15th, then $t_{2,15,1}=0.1$
means player 2 spent 0.1 hour on December 15th between 0AM and 1AM.
Similarly, $t_{2,1,2}=0.4$ means player 2 spent 0.4 hour on December
1st between 1AM and 2AM. A different player may have a different latest
playing date. For example, if player 3's latest playing date is November
20th, then $t_{3,15,4}=0.7$ means player 3 spent 0.7 hour on November
20th between 3AM and 4AM. Formally, $d$ indexes the $n+1-d$-th day
to the latest playing date.

We partition $n$ days into $m$ periods of equal days. For example,
if $n=15$, $m=5$, then each period has 3 days and the first period
includes days in the set $D_{1}=\{1,2,3\}$ and the second period
includes days in the set $D_{2}=\{4,5,6\}$. Formally, the $k$-th
period includes days in
\begin{eqnarray*}
D_{k} & = & \left\{ (k-1)\frac{n}{m}+i:i=1,2,...,\frac{n}{m}\right\} ,\ k=1,2,...,m.
\end{eqnarray*}

Based on the latest playing times, we will calculate the empirical
probability distributions related to the in-game time spent by a player.

\subsection{Daily Playtime Distribution}

We first consider the total in-game time spent on each day and how
each player distributes his in-game time over different days of a
period. To this end, we define the individual (empirical) probability
of the playtime for player $u$ on day $d$ within period $k$ as%
\begin{align*}
p_{\text{ind}}(d|u,k) & =\frac{\sum_{r=1}^{24}t_{u,d,r}}{\sum_{w\in D_{k}}\sum_{r=1}^{24}t_{u,w,r}}.
\end{align*}
For instance, consider the latest $n=15$ days of playing the game
for a certain player, with $m=5$ periods of 3 days. Assume that during
the first period player $u$ spends $5$ hour, $6$ hours, and $8$
hours on the 1st, $2$nd, and $3$rd days, respectively. Then within
the first period, the probabilities of the daily playtime of this
player over the 3 days are $p_{\text{ind}}(1|u,1)=5/(5+6+8)$, $p_{\text{ind}}(2|u,1)=6/(5+6+8)$,
and $p_{\text{ind}}(3|u,1)=8/(5+6+8)$.

In addition, in order to capture the daily playtime distribution of
the community of churners in set $U$, we introduce a ``global''
probability of the total daily playtime of all the churners on the
$d$-th day within the $k$-th period as
\begin{eqnarray*}
p_{\text{global}}^{\text{churner}}(d|k) & = & \frac{\sum_{u\in U}\sum_{r=1}^{24}t_{u,d,r}}{\sum_{u\in U}\sum_{w\in D_{k}}\sum_{r=1}^{24}t_{u,w,r}}.
\end{eqnarray*}

\subsection{Hourly Playtime Distribution}

We next consider the in-game time spent in each hour and how it is
distributed over different days of a period. We define the empirical
probability of the playtime for player $u$ in hour $r$ on day $d$
within period $k$ as
\[
P_{\text{ind}}(d|u,k,r)=\frac{t_{u,d,r}}{\sum_{w\in D_{k}}t_{u,w,r}}.
\]
For instance, consider the same example in the last subsection, and
assume that during the first period player $u$ spent $0.1$ hour,
$0.2$ hour, and $0.3$ hour in the hour from 8:00 - 9:00 on the $1$st,
$2$nd, and $3$rd days, respectively. Then within the first period,
the probabilities of the playtime in the hour from 8:00 - 9:00 of
this player over the first period are $P_{\text{ind}}(1|u,1,9)=0.1/(0.1+0.2+0.3)$,
$P_{\text{ind}}(2|u,1,9)=0.2/(0.1+0.2+0.3)$, and $P_{\text{ind}}(3|u,1,9)=0.3/(0.1+0.2+0.3)$.

Similarly, in order to capture the hourly playtime distribution of
the churner community $U$, we introduce a global probability of the
playtime of all the churners in hour $r$ on the $d$-th day within
the $k$-th period as
\begin{eqnarray*}
p_{\text{global}}^{\text{churner}}(d|k,r) & = & \frac{\sum_{u\in U}t_{u,d,r}}{\sum_{u\in U}\sum_{w\in D{}_{k}}t_{u,w,r}}.
\end{eqnarray*}
\\

The afore-introduced players' playtime distributions will be the basis
to extract new features for churn prediction.

\section{Time Spending Regularity: Churners Versus Non-churners\label{sec:Mining-Player-Time}}

Since we aim to predict churn in this paper, for each game, the data
set is equally divided into a training data set and a test data set.
In this section, we examine the playtime patterns of churners and
non-churners from the training data set at different timescales for
each game, with the aim of identifying possible differentiators between
churners and non-churners.

\subsection{Entropy and Playtime Pattern \label{subsec:Period-Variation-Signals}}

Based on the players' playtime distributions introduced in Section
\ref{sec:Representing-Player-Time} and motivated by \cite{Rowe_2016ATKDD_MiningUserDevelopmentSignalsOnlineCommunityChurnerDetection-1},
we use the notion of \textit{entropy} from information theory as the
metric to characterize variance and change in playtime \cite{Thomas_2006_ElementsInformationTheory2Ed}.
Given a probability distribution $p(\cdot)$, its entropy is defined
as
\[
H(p)=\sum_{x}p(x)\log\frac{1}{p(x)}.
\]
A higher entropy means a more even distribution and more regular playtime
pattern, and a smaller entropy implies a less even distribution and
more irregular/casual playtime pattern.

For each game and the data set described in Section \ref{sec:Free-Online-Game},
we consider the latest $n=15$ days playing time, which is the same
as \cite{Liu_20192ICGC_MiningPlayerIngameTimeSpendingRegularityChurnPredictionFreeOnlineGames}.
But we consider multiple divisions of playtime to periods to examine players' playtime patterns
where \cite{Liu_20192ICGC_MiningPlayerIngameTimeSpendingRegularityChurnPredictionFreeOnlineGames}
only considers a single division of playtime. In this paper, we divide the playtime
into $m=2,3,5$ periods, where the corresponding periods have $7,5,3$
days, respectively. We calculate the corresponding playtime distributions
and entropies for each player of different periods.

\begin{figure*}[t]
\hfill{}\subfloat{\centering{}\includegraphics[clip,width=0.8\paperwidth]{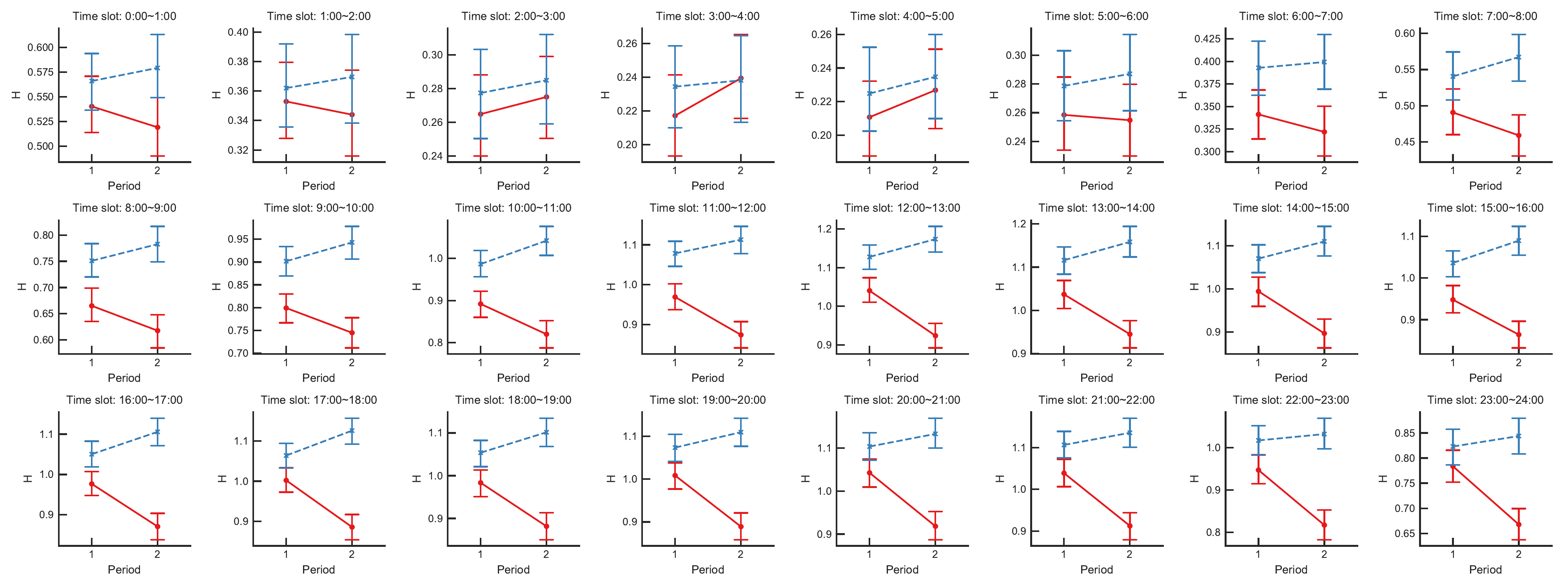}}\hfill{}

\caption{\label{fig:3}The mean entropies of the distributions of hourly time
spending of \textit{Thirty-six Stratagems (TS)} players where the
playing time is divided by $m=2$ periods. The blue dashed line shows
the non-churners, while the red solid line shows the churners.}

\hfill{}\subfloat{\includegraphics[clip,width=0.8\paperwidth]{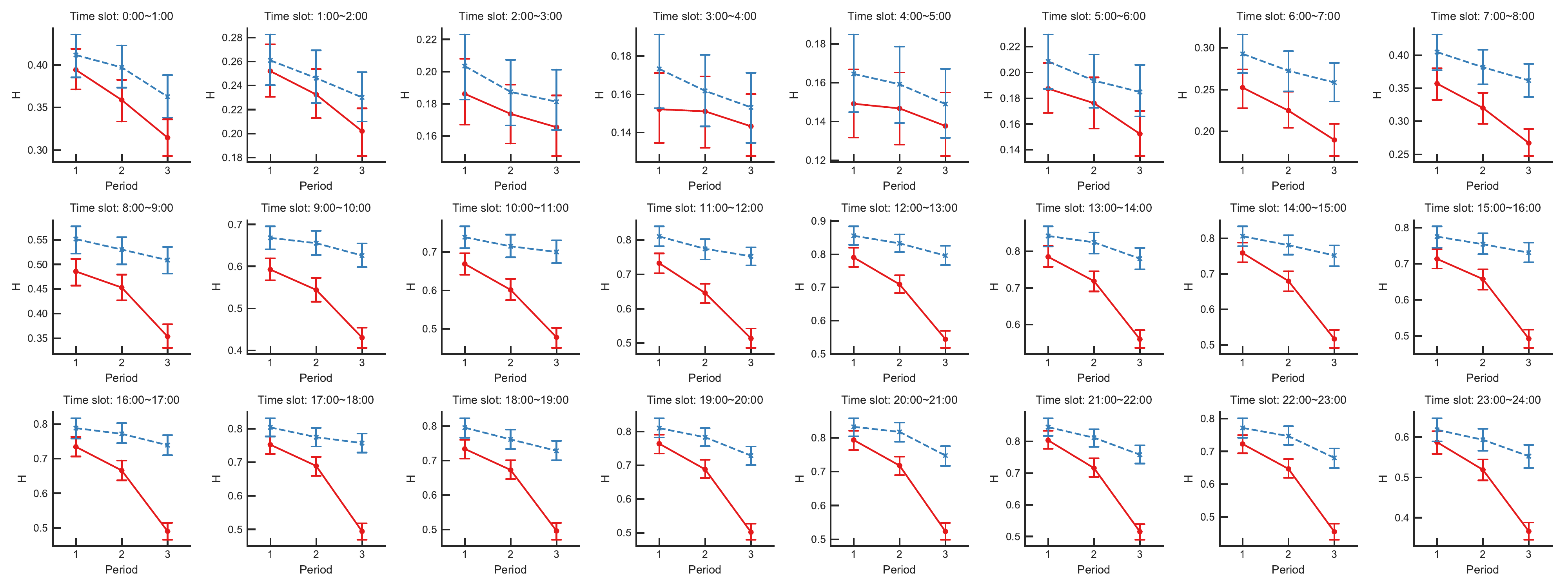}}\hfill{}

\caption{\label{fig:DailyCrossEntropyC-1}The mean entropies of the distributions
of hourly time spending of \textit{Thirty-six Stratagems (TS)} players
where the playing time is divided by $m=3$ periods. The blue dashed
line shows the non-churners, while the red solid line shows the churners.}
\hfill{}\subfloat{\includegraphics[clip,width=0.8\paperwidth]{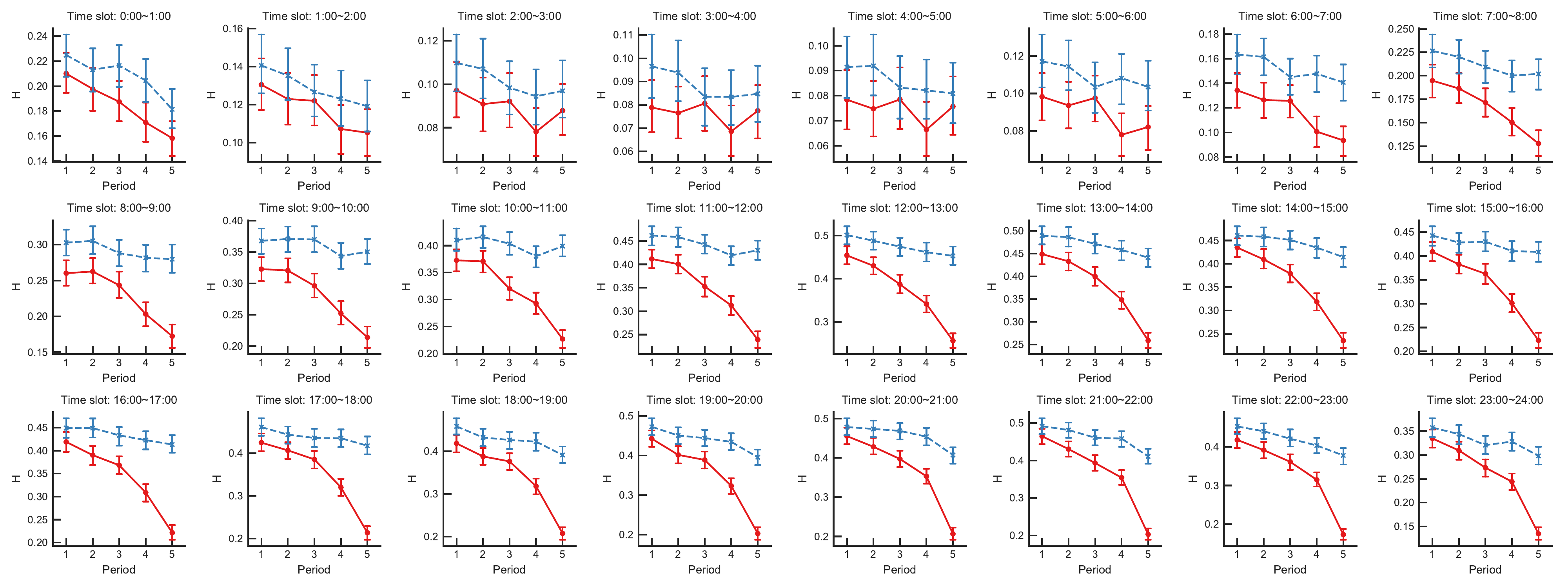}}\hfill{}

\caption{\label{fig:DailyCrossEntropyNC-1}The mean entropies of the distributions
of hourly time spending of \textit{Thirty-six Stratagems (TS)} players
where the playing time is divided by $m=5$ periods. The blue dashed
line shows the non-churners, while the red solid line shows the churners.}
\end{figure*}

\begin{figure*}[t]
\hfill{}\subfloat{\centering{}\includegraphics[clip,width=0.8\paperwidth]{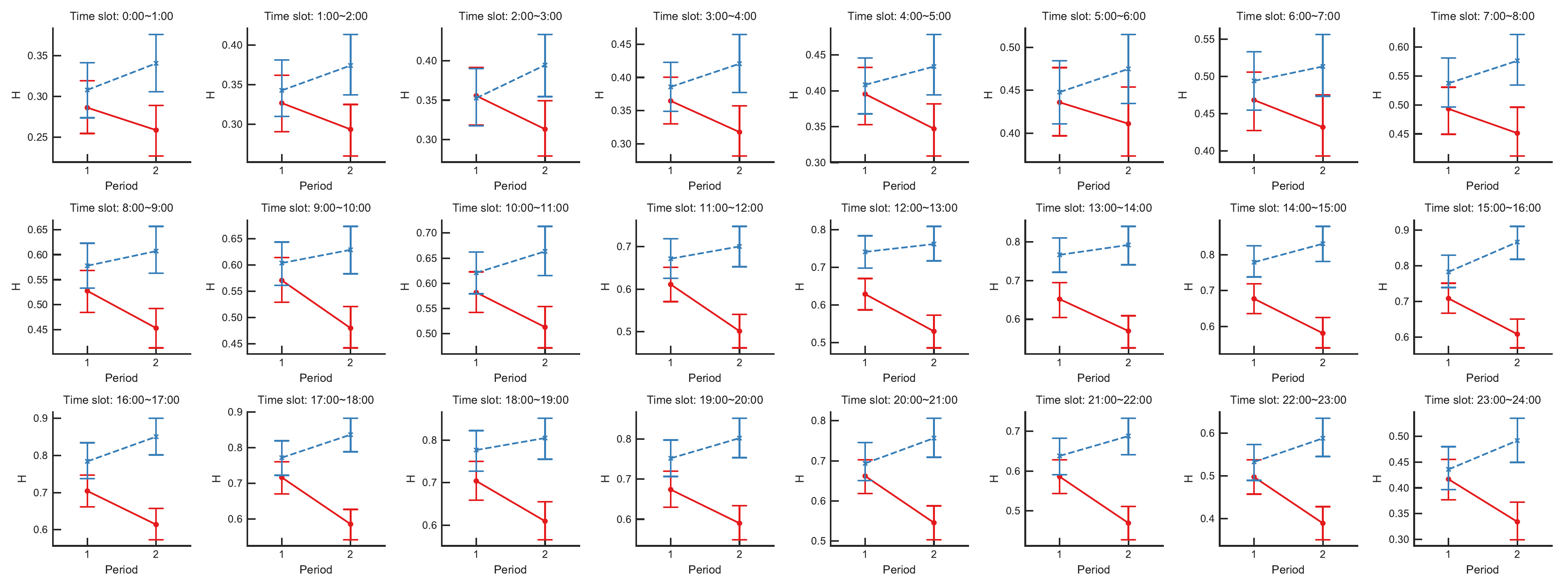}}\hfill{}

\caption{\label{fig:DailyEntropy-1-1}The mean entropies of the distributions
of hourly time spending of \textit{Game of Thrones Winter is Coming
(GOT)} players where the playing time is divided by $m=2$ periods.
The blue dashed line shows the non-churners, while the red solid line
shows the churners.}

\hfill{}\subfloat{\includegraphics[clip,width=0.8\paperwidth]{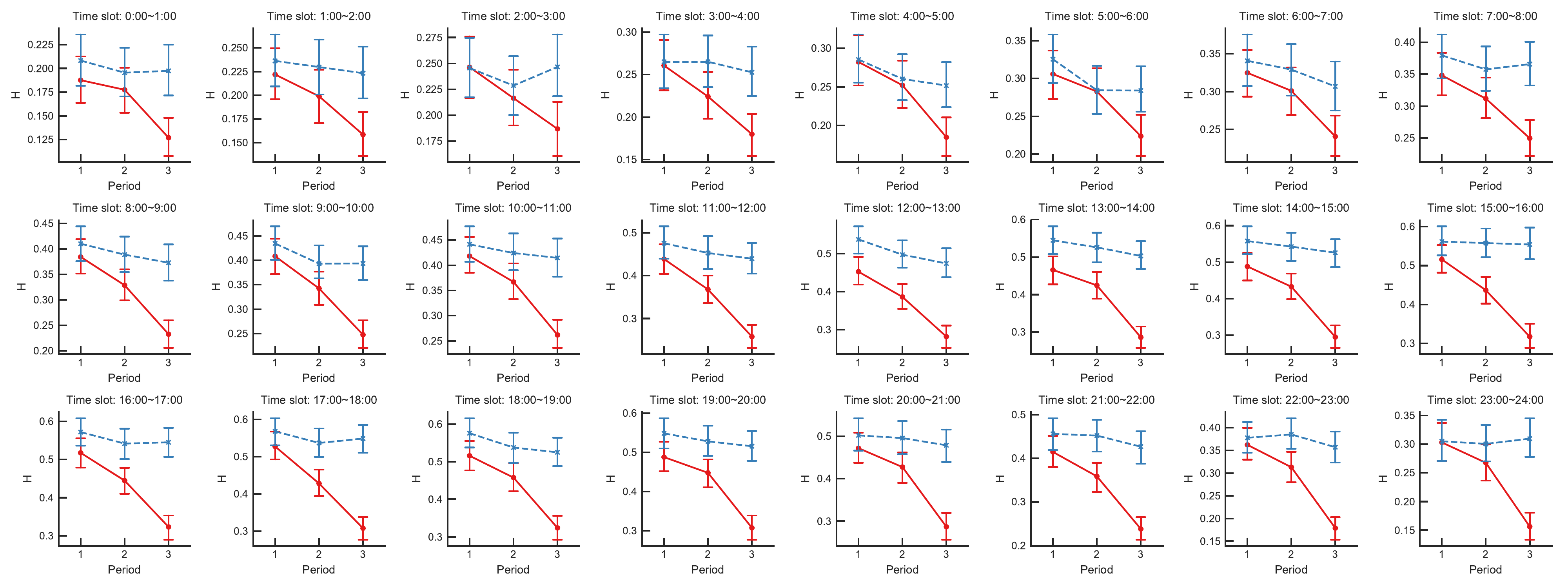}}\hfill{}

\caption{\label{fig:DailyCrossEntropyC-1-1}The mean entropies of the distributions
of hourly time spending of \textit{Game of Thrones Winter is Coming
(GOT)} players where the playing time is divided by $m=3$ periods.
The blue dashed line shows the non-churners, while the red solid line
shows the churners.}
\hfill{}\subfloat{\includegraphics[clip,width=0.8\paperwidth]{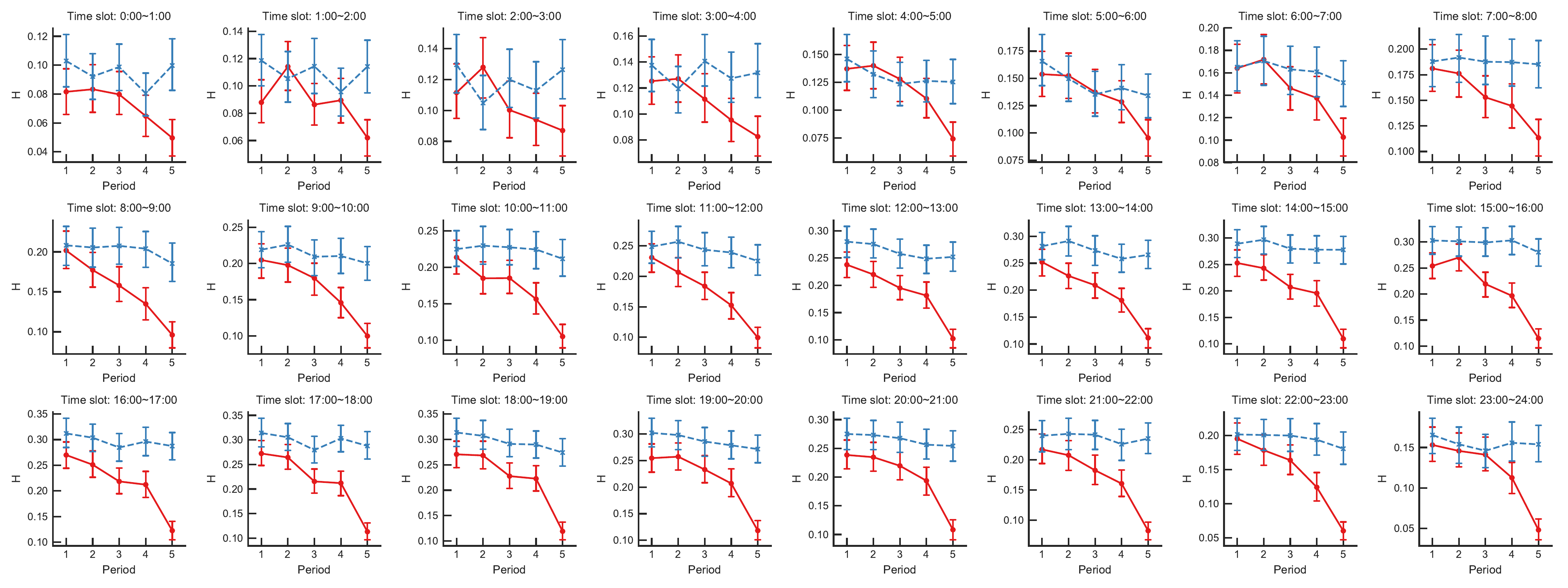}}\hfill{}

\caption{\label{fig:DailyCrossEntropyNC-1-1}The mean entropies of the distributions
of hourly time spending of \textit{Game of Thrones Winter is Coming
s (GOT)} players where the playing time is divided by $m=5$ periods.
The blue dashed line shows the non-churners, while the red solid line
shows the churners.}
\end{figure*}

\begin{figure*}[t]
\hfill{}\subfloat{\centering{}\includegraphics[clip,width=0.8\paperwidth]{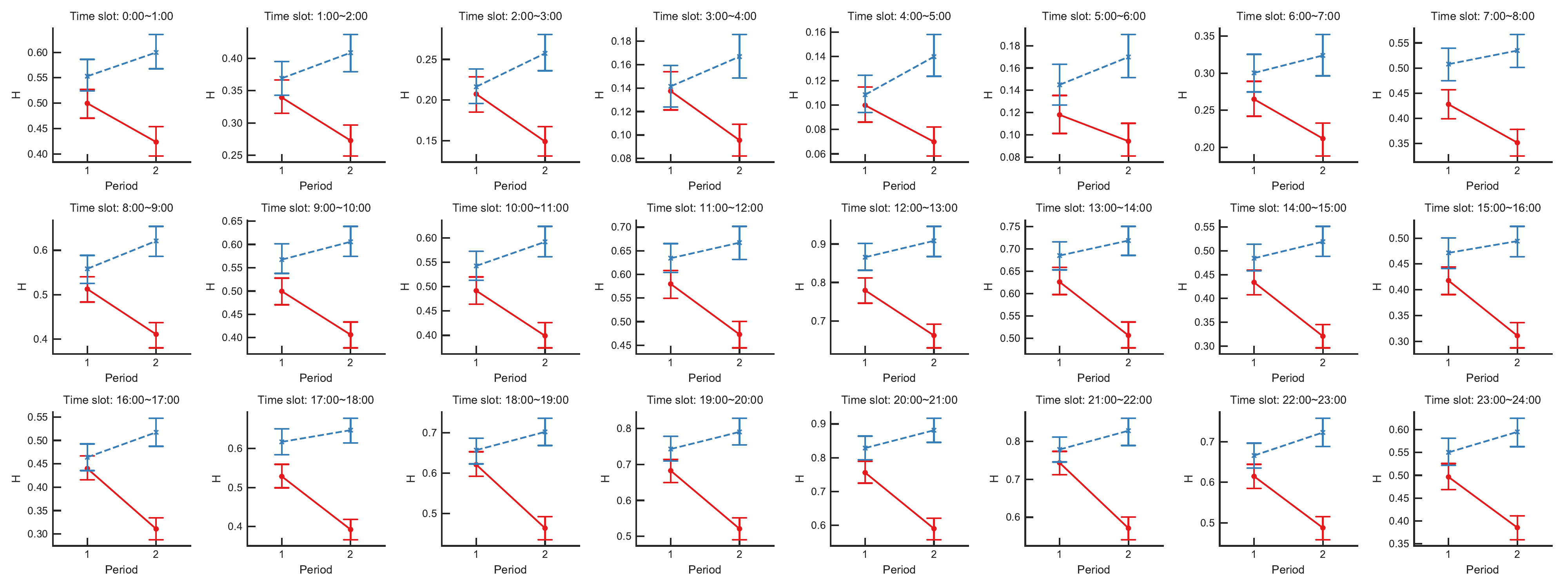}}\hfill{}

\caption{\label{fig:DailyEntropy-1-2}The mean entropies of the distributions
of hourly time spending of \textit{Thirty-six Stratagems} \textit{Mobile
(TSM)} players where the playing time is divided by $m=2$ periods.
The blue dashed line shows the non-churners, while the red solid line
shows the churners.}

\hfill{}\subfloat{\includegraphics[clip,width=0.8\paperwidth]{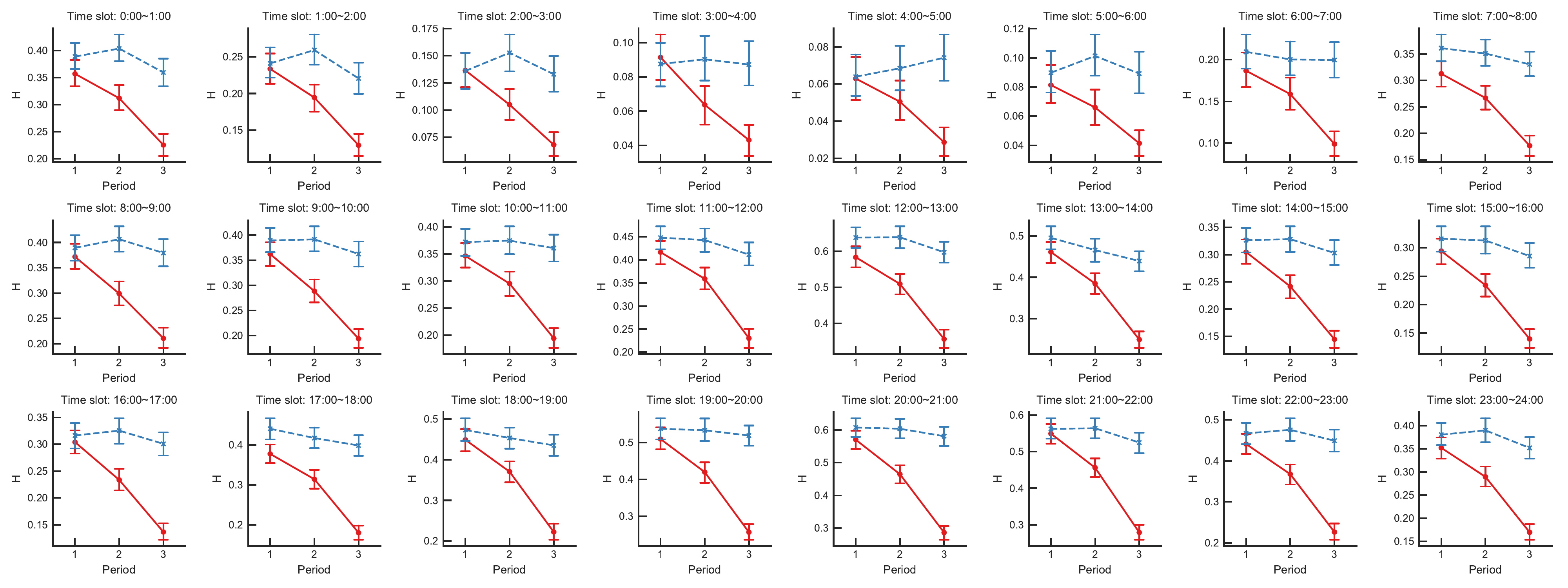}}\hfill{}

\caption{\label{fig:DailyCrossEntropyC-1-2}The mean entropies of the distributions
of hourly time spending of \textit{Thirty-six Stratagems} \textit{Mobile
(TSM)} players where the playing time is divided by $m=3$ periods.
The blue dashed line shows the non-churners, while the red solid line
shows the churners.}
\hfill{}\subfloat{\includegraphics[clip,width=0.8\paperwidth]{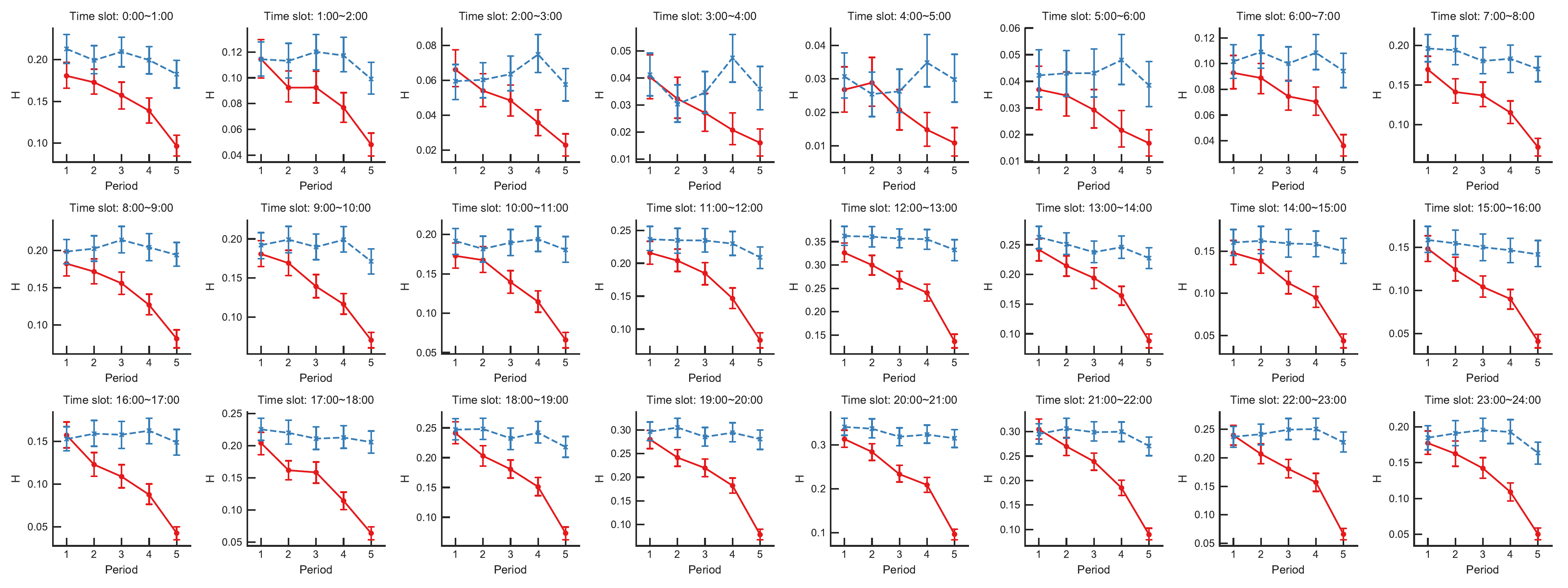}}\hfill{}

\caption{\label{fig:DailyCrossEntropyNC-1-2}The mean entropies of the distributions
of hourly time spending of \textit{Thirty-six Stratagems} \textit{Mobile
(TSM)} players where the playing time is divided by $m=5$ periods.
The blue dashed line shows the non-churners, while the red solid line
shows the churners.}
\end{figure*}

\begin{figure*}[t]
\hfill{}\subfloat{\centering{}\includegraphics[clip,width=0.8\paperwidth]{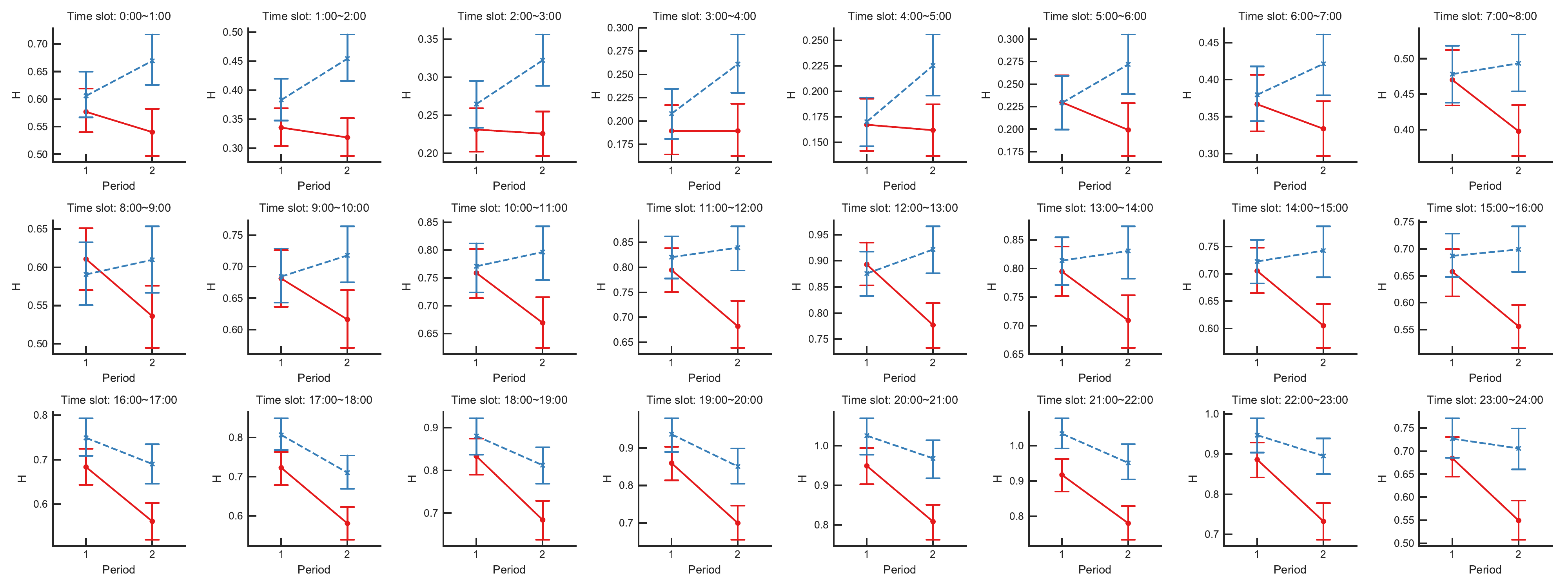}}\hfill{}

\caption{\label{fig:12}The mean entropies of the distributions of hourly time
spending of \textit{Womanland in Journey to the West (WJW)} players
where the playing time is divided by $m=2$ periods. The blue dashed
line shows the non-churners, while the red solid line shows the churners.}

\hfill{}\subfloat{\includegraphics[clip,width=0.8\paperwidth]{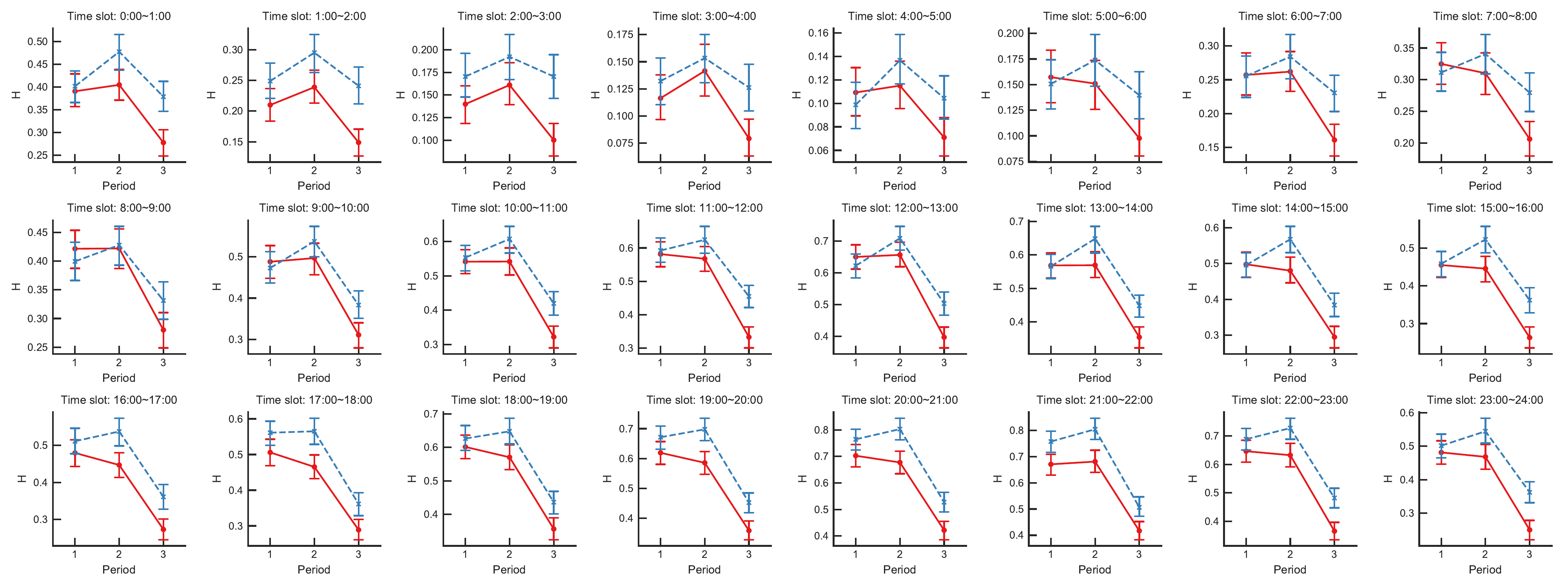}}\hfill{}

\caption{\label{fig:DailyCrossEntropyC-1-3}The mean entropies of the distributions
of hourly time spending of \textit{Womanland in Journey to the West
(WJW)} players where the playing time is divided by $m=3$ periods.
The blue dashed line shows the non-churners, while the red solid line
shows the churners.}
\hfill{}\subfloat{\includegraphics[clip,width=0.8\paperwidth]{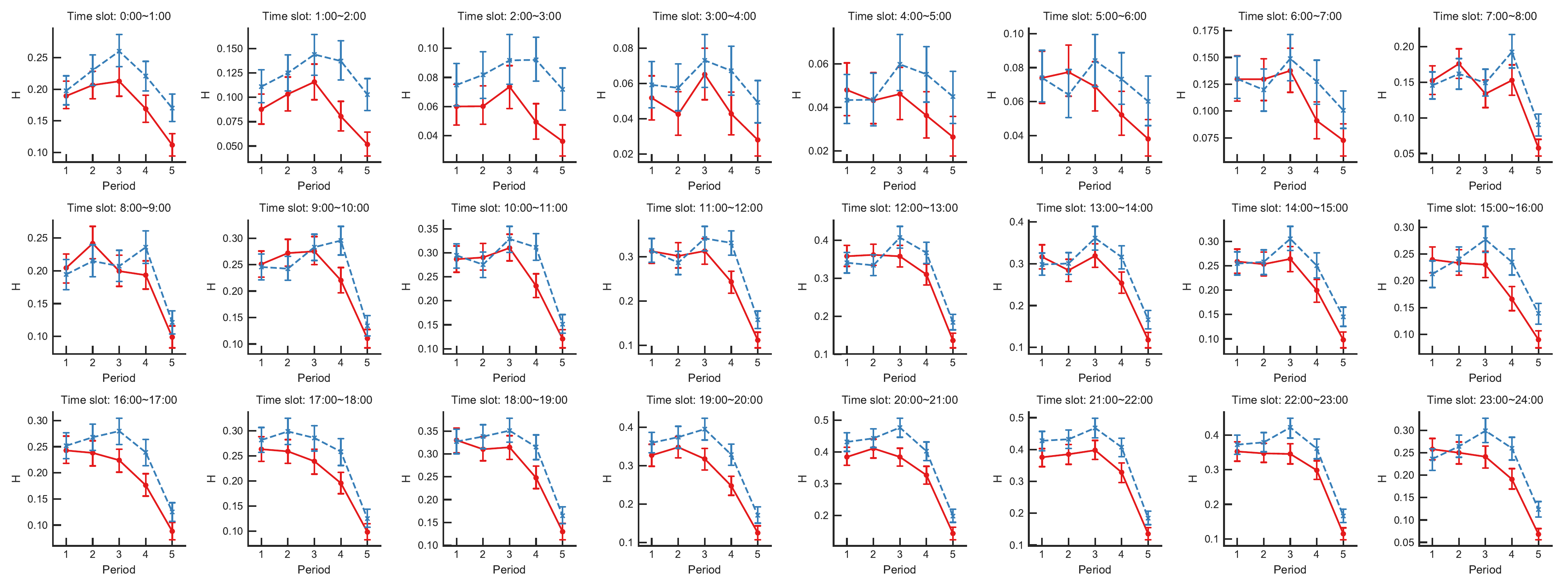}}\hfill{}

\caption{\label{fig:14}The mean entropies of the distributions of hourly time
spending of \textit{Womanland in Journey to the West (WJW)} players
where the playing time is divided by $m=5$ periods. The blue dashed
line shows the non-churners, while the red solid line shows the churners.}
\end{figure*}

\begin{figure*}[t]
\hfill{}\subfloat{\centering{}\includegraphics[clip,width=0.8\paperwidth]{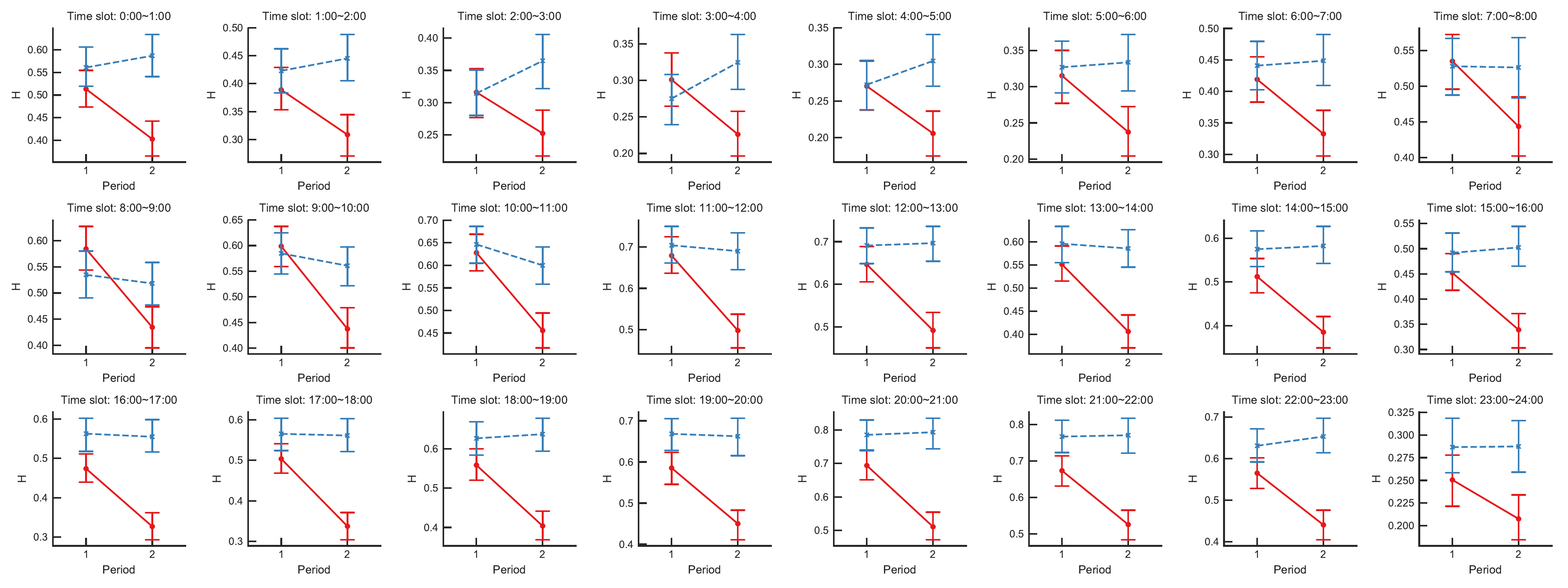}}\hfill{}

\caption{\label{fig:DailyEntropy-1-4}The mean entropies of the distributions
of hourly time spending of \textit{Era of Angels (EOA)} players where
the playing time is divided by $m=2$ periods. The blue dashed line
shows the non-churners, while the red solid line shows the churners.}

\hfill{}\subfloat{\includegraphics[clip,width=0.8\paperwidth]{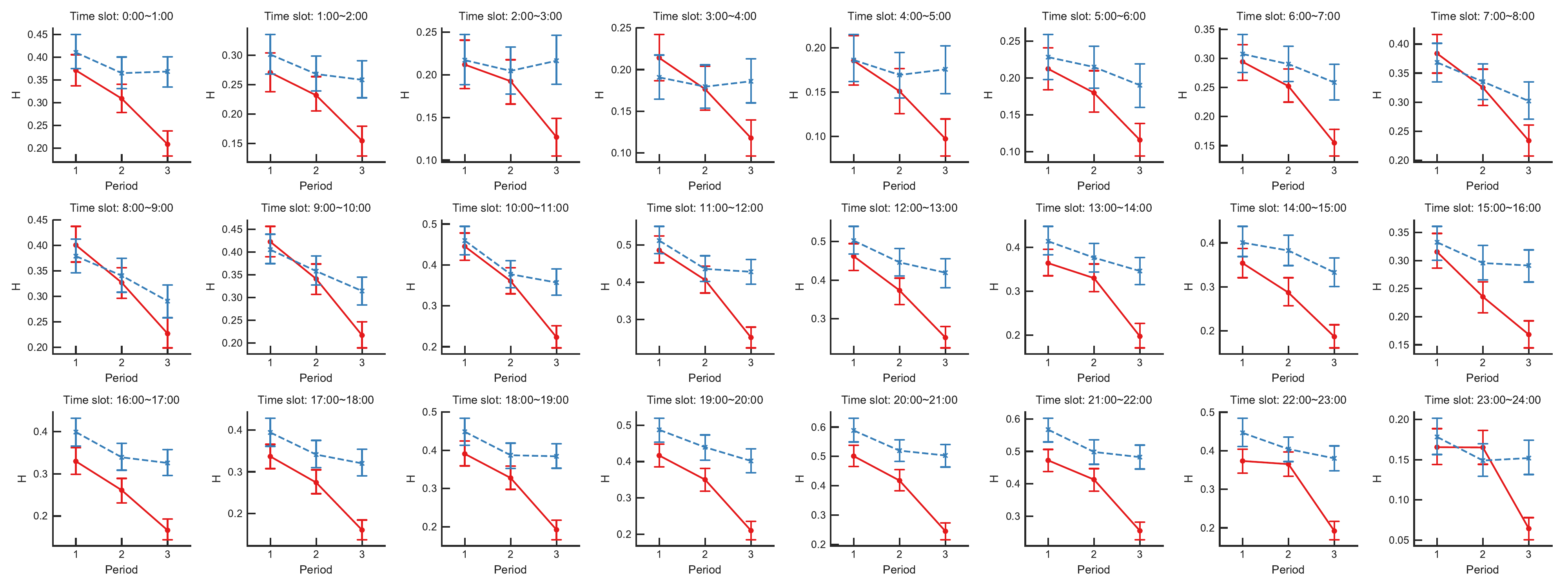}}\hfill{}

\caption{\label{fig:DailyCrossEntropyC-1-4}The mean entropies of the distributions
of hourly time spending of \textit{Era of Angels (EOA)} players where
the playing time is divided by $m=3$ periods. The blue dashed line
shows the non-churners, while the red solid line shows the churners.}
\hfill{}\subfloat{\includegraphics[clip,width=0.8\paperwidth]{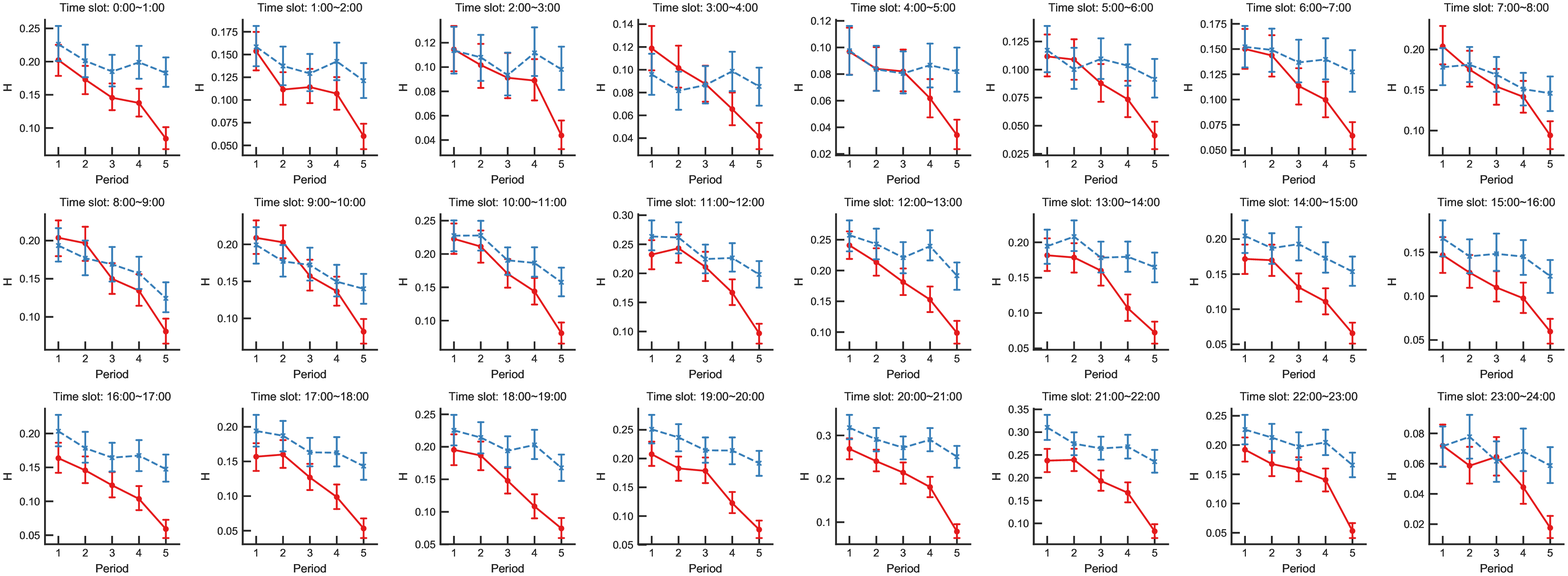}}\hfill{}

\caption{\label{fig:DailyCrossEntropyNC-1-4}The mean entropies of the distributions
of hourly time spending of \textit{Era of Angels (EOA)} players where
the playing time is divided by $m=5$ periods. The blue dashed line
shows the non-churners, while the red solid line shows the churners.}
\end{figure*}

\begin{figure*}[t]
\hfill{}\subfloat{\centering{}\includegraphics[clip,width=0.8\paperwidth]{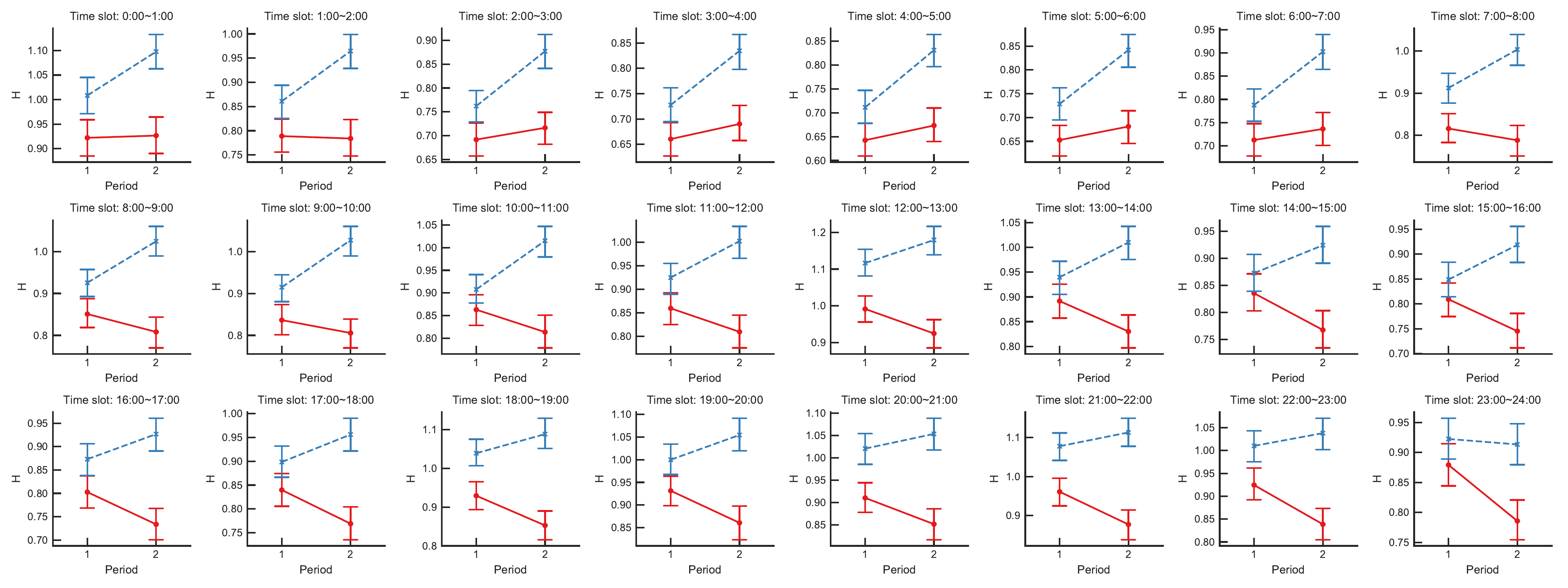}}\hfill{}

\caption{\label{fig:DailyEntropy-1-5}The mean entropies of the distributions
of hourly time spending of \textit{League of Angels II (LOA II)} players
where the playing time is divided by $m=2$ periods. The blue dashed
line shows the non-churners, while the red solid line shows the churners.}

\hfill{}\subfloat{\includegraphics[clip,width=0.8\paperwidth]{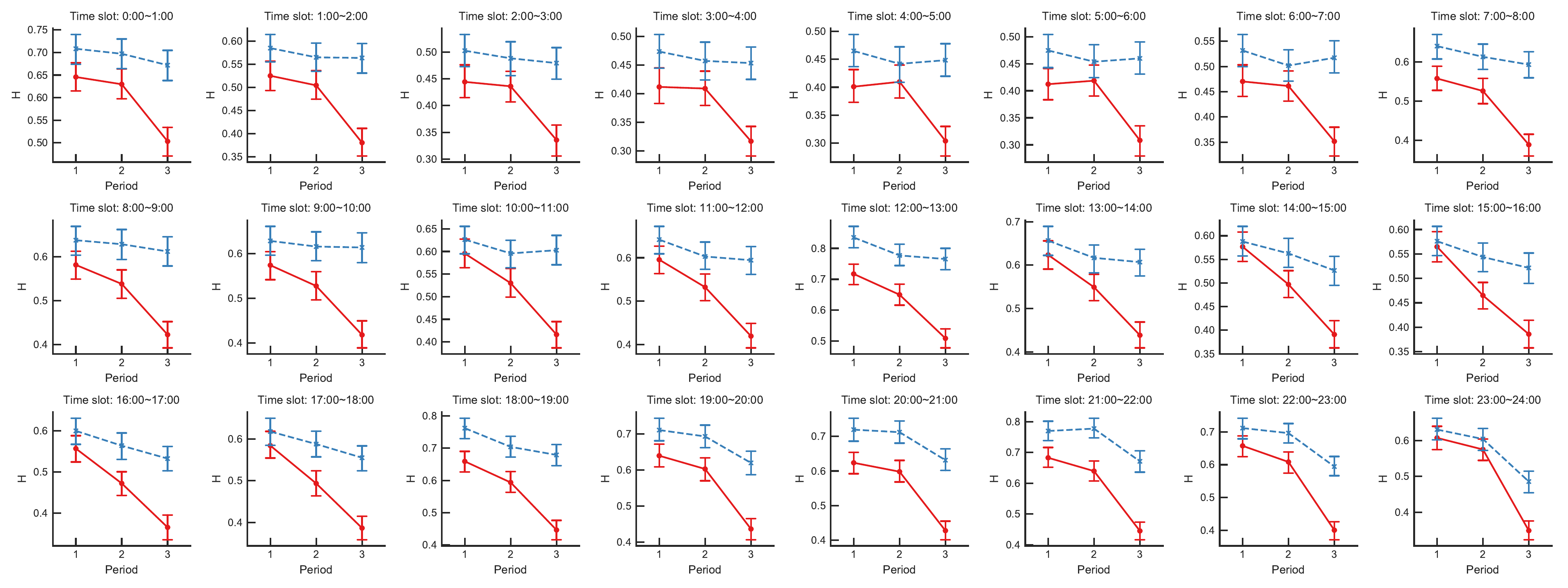}}\hfill{}

\caption{\label{fig:DailyCrossEntropyC-1-5}The mean entropies of the distributions
of hourly time spending of \textit{League of Angels II (LOA II)} players
where the playing time is divided by $m=3$ periods. The blue dashed
line shows the non-churners, while the red solid line shows the churners.}
\hfill{}\subfloat{\includegraphics[clip,width=0.8\paperwidth]{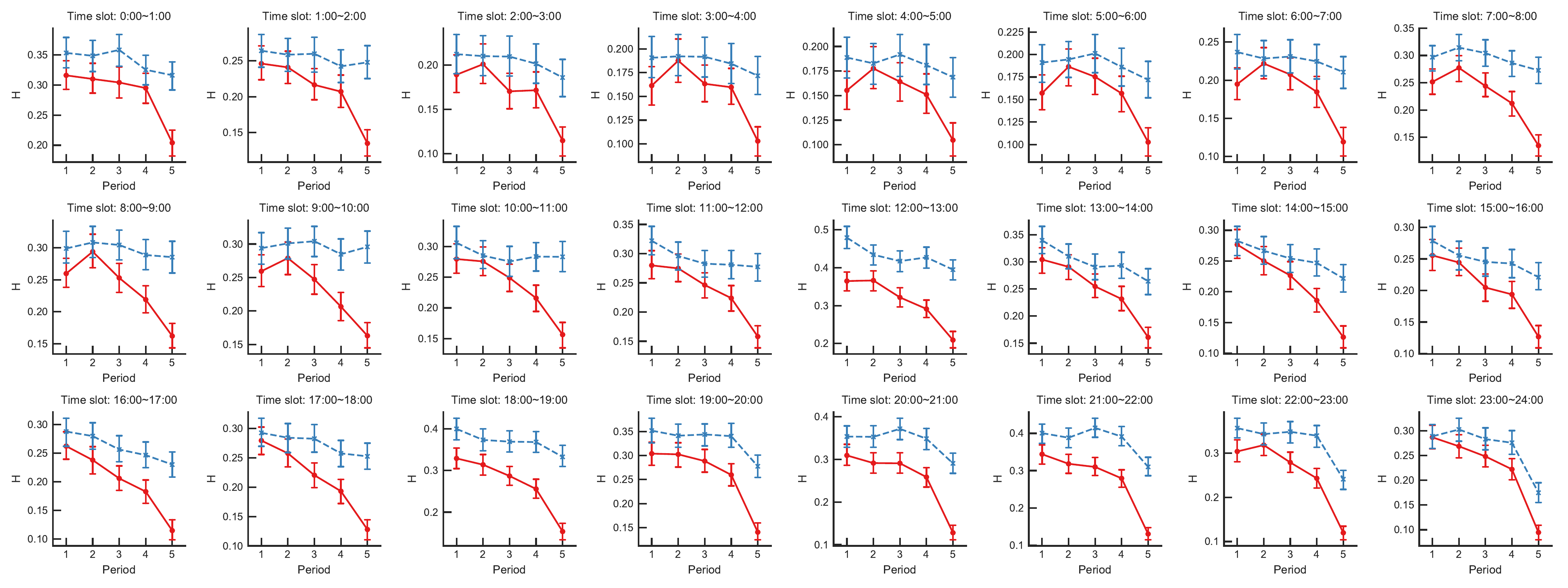}}\hfill{}

\caption{\label{fig:20}The mean entropies of the distributions of hourly time
spending of \textit{League of Angels II (LOA II)} players where the
playing time is divided by $m=5$ periods. The blue dashed line shows
the non-churners, while the red solid line shows the churners.}
\end{figure*}

The mean entropies and $95\%$ confidence intervals of churners and
non-churners in different periods are shown in Appendix \ref{sec:The-Entropy-Distributions}
(Fig. \ref{fig:DailyEntropy}, Fig. \ref{fig:3} - Fig. \ref{fig:20}).
Fig. \ref{fig:DailyEntropy} illustrates the entropy distributions
of churners and non-churners in those games on a granularity of each
day for all aforementioned games, while Fig. \ref{fig:3} - Fig. \ref{fig:20}
illustrate the entropy distributions on a granularity of each hour.

We observe that non-churners have a higher mean value of entropy than
churners for any selected number of periods $m$, in both daily and
hourly playtime distributions and in those games. This implies that
non-churners have much more regular playtime patterns than churners.
Moreover, the entropies of non-churners exhibit a smaller decrease
as time moves on, compared to churners. This implies that non-churners
have a more regular playtime pattern than churners across different
timescales, while churners spend their in-game time more and more
casually as time moves on.

\begin{figure*}[t]
\hfill{}\subfloat{\centering{}\includegraphics[clip,width=0.8\paperwidth]{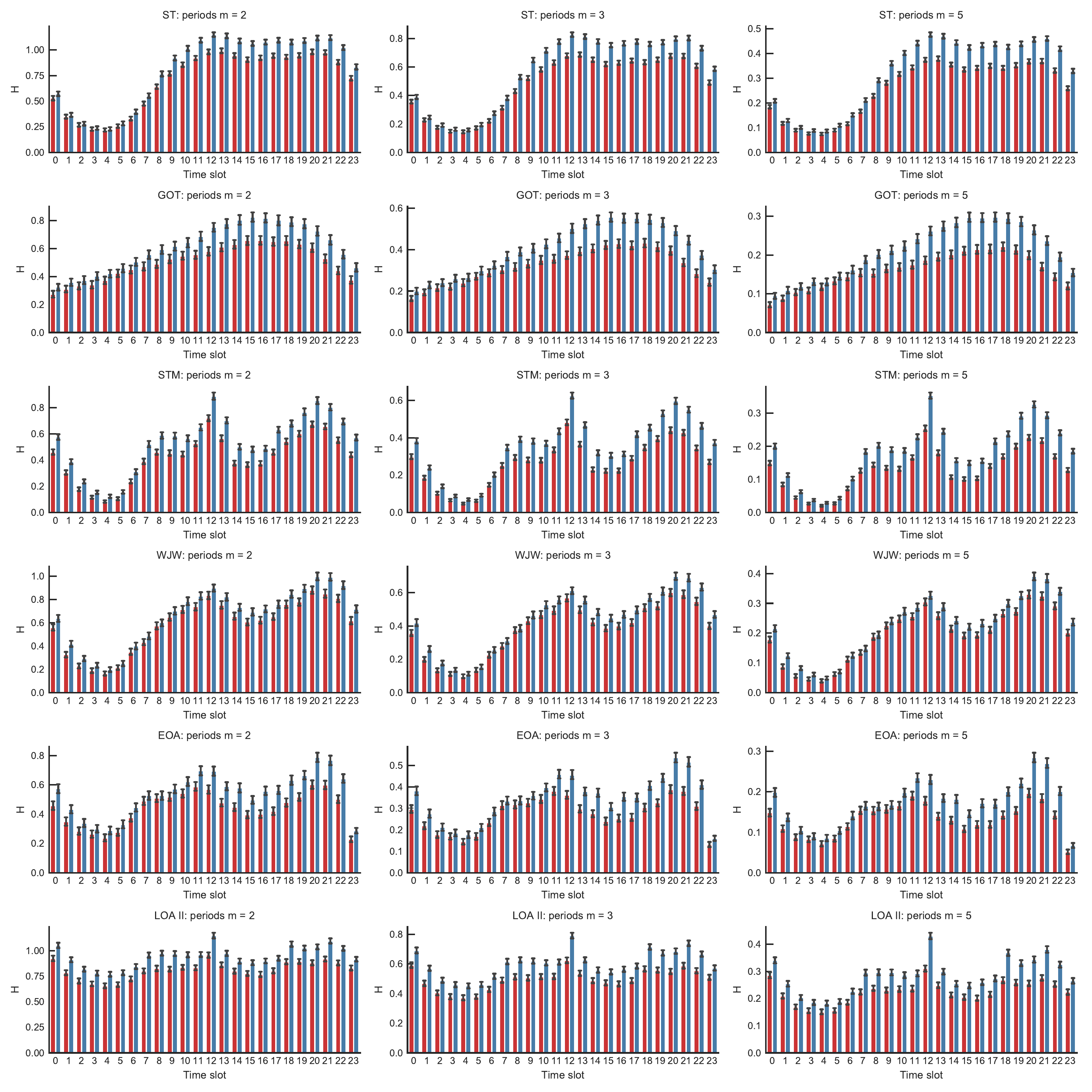}}\hfill{}

\caption{\label{fig:-2}The mean entropies of the distributions of hourly time
spending of aforementioned games' players in different time slot (from
0:00 - 1:00 to 23:00 - 24:00, indexed from 0 - 23). The blue bar shows
the non-churners, while the red bar shows the churners.}
\end{figure*}

A further look at the hourly entropies in Appendix \ref{sec:The-Entropy-Distributions}
(Fig. \ref{fig:3} - Fig. \ref{fig:20} and Fig. \ref{fig:-2}) shows
that the above mentioned difference between churners and non-churners
is more significant in the hours from early morning to late night
and less significant from late night to early morning. In Fig. \ref{fig:-2},
we can see that the playtime entropy/regularity%
{} is higher from early morning to late night and lower from late night
to early morning. This is consistent with the fact that late night
to early morning is the most common sleep time,%
{} and it is hard for the majority of players (churners or non-churners)
to maintain a regular playtime pattern during that period of time.

\subsection{Cross-entropy and Correlation with Aggregate Patterns\label{subsec:Gaming-Community-Comparison}}

In our previous work \cite{Liu_20192ICGC_MiningPlayerIngameTimeSpendingRegularityChurnPredictionFreeOnlineGames},
we examined how each player compares to the entire game community
that he/she is playing with over the same corresponding playing period
using the notion of \textit{cross-entropy} from information theory.
However, the examination results may vary if the proportion of churners/non-churners
changes in the dataset.

In this paper, instead of examining how each player compares to the
entire game community, we now examine how each player compares to
the churner community. Given an individual player's daily playtime
distribution $p_{\text{ind}}(d|u,k)$ and the global daily distribution
$p_{\text{global}}^{\text{churner}}(d|k)$ of the churners , the cross-entropy between these two distributions is defined as%
\begin{align*}
 & H(p_{\text{ind}}(\cdot|u,k),p_{\text{global}}^{\text{churner}}(\cdot|k))=\\
 & \ \ \ \ \ \ \ \ \ \sum_{d}p_{\text{ind}}(d|u,k)\log\frac{1}{p_{\text{global}}^{\text{churner}}(d|k)}.
\end{align*}
{} Similarly, for hourly distribution, the cross-entropy is

\begin{align*}
 & H(p_{\text{ind}}(\cdot|u,k,r),p_{\text{global}}^{\text{churner}}(\cdot|k,r))=\\
 & \ \ \ \ \ \ \ \ \ \sum_{d}p_{\text{ind}}(d|u,k,r)\log\frac{1}{p_{\text{global}}^{\text{churner}}(d|k,r)}.
\end{align*}
For example, if the cross-entropy is smaller, it implies that the
playtime pattern of the player is more similar to that of churners'
community.

\begin{figure*}[t]
\hfill{}\subfloat{\centering{}\includegraphics[clip,width=0.8\paperwidth]{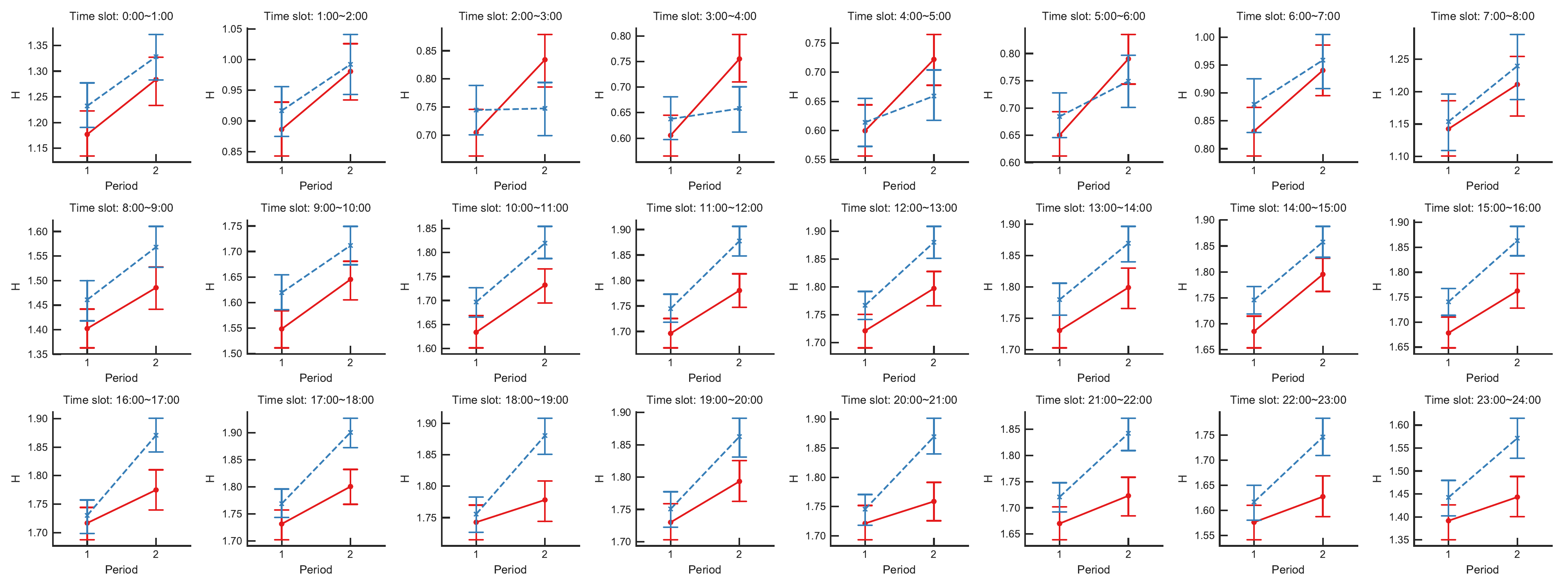}}\hfill{}

\caption{\label{fig:21}The mean cross-entropies of the distributions of hourly
time spending between players and churner community in \textit{Thirty-six
Stratagems (TS)} where the playing time is divided by $m=2$ periods.
The blue dashed line shows the non-churners, while the red solid line
shows the churners.}

\hfill{}\subfloat{\includegraphics[clip,width=0.8\paperwidth]{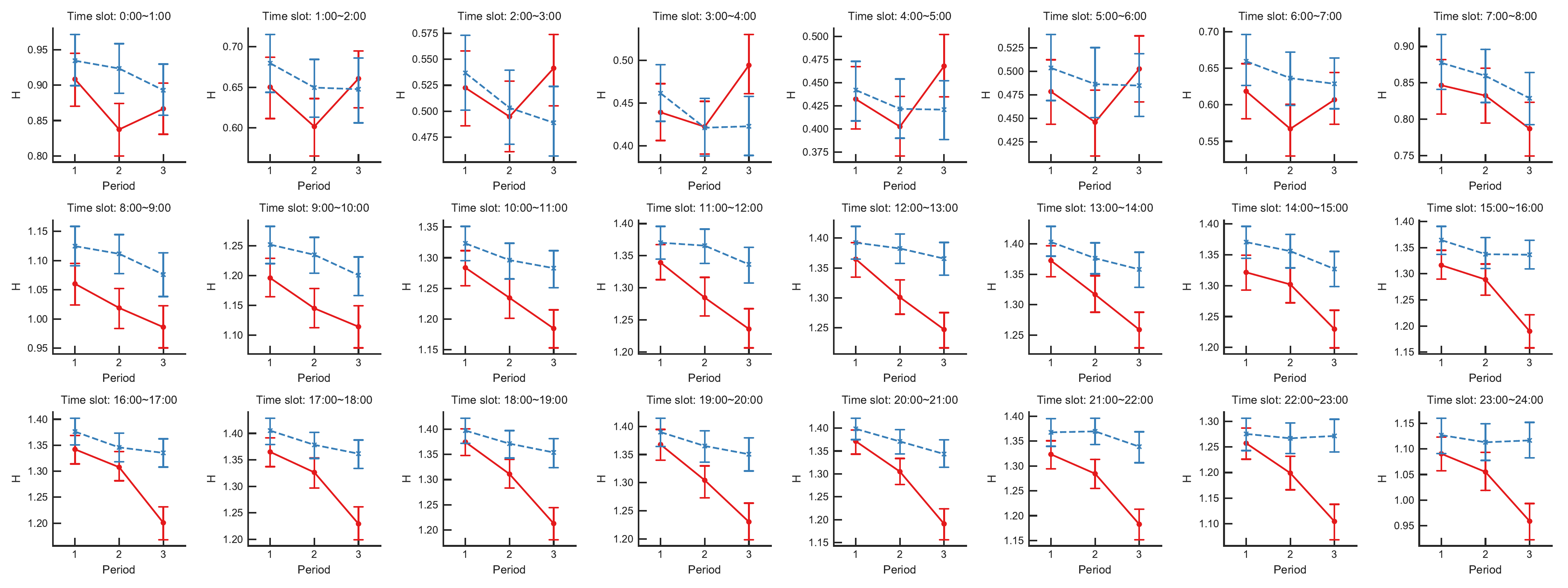}}\hfill{}

\caption{\label{fig:DailyCrossEntropyC-1-6}The mean cross-entropies of the
distributions of hourly time spending between players and churner
community in \textit{Thirty-six Stratagems (TS)} where the playing
time is divided by $m=3$ periods. The blue dashed line shows the
non-churners, while the red solid line shows the churners.}
\hfill{}\subfloat{\includegraphics[clip,width=0.8\paperwidth]{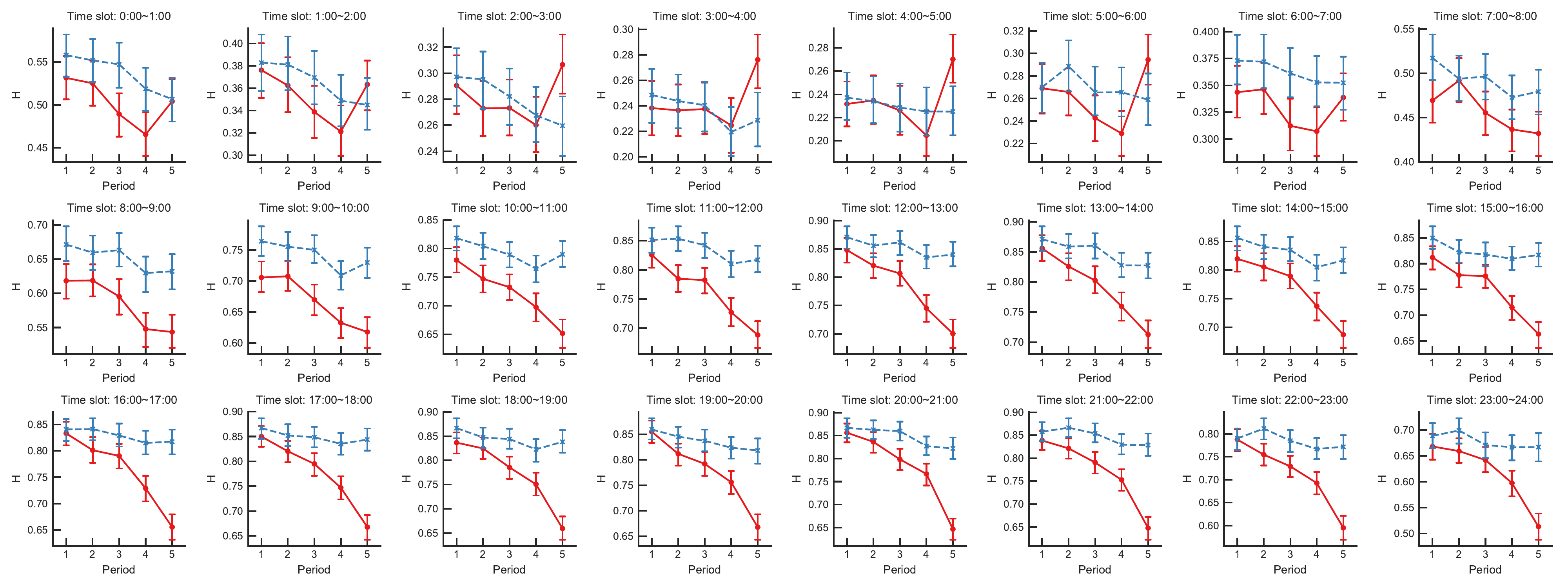}}\hfill{}

\caption{\label{fig:DailyCrossEntropyNC-1-6}The mean cross-entropies of the
distributions of hourly time spending between players and churner
community in \textit{Thirty-six Stratagems (TS)} where the playing
time is divided by $m=5$ periods. The blue dashed line shows the
non-churners, while the red solid line shows the churners.}
\end{figure*}

\begin{figure*}[t]
\hfill{}\subfloat{\centering{}\includegraphics[clip,width=0.8\paperwidth]{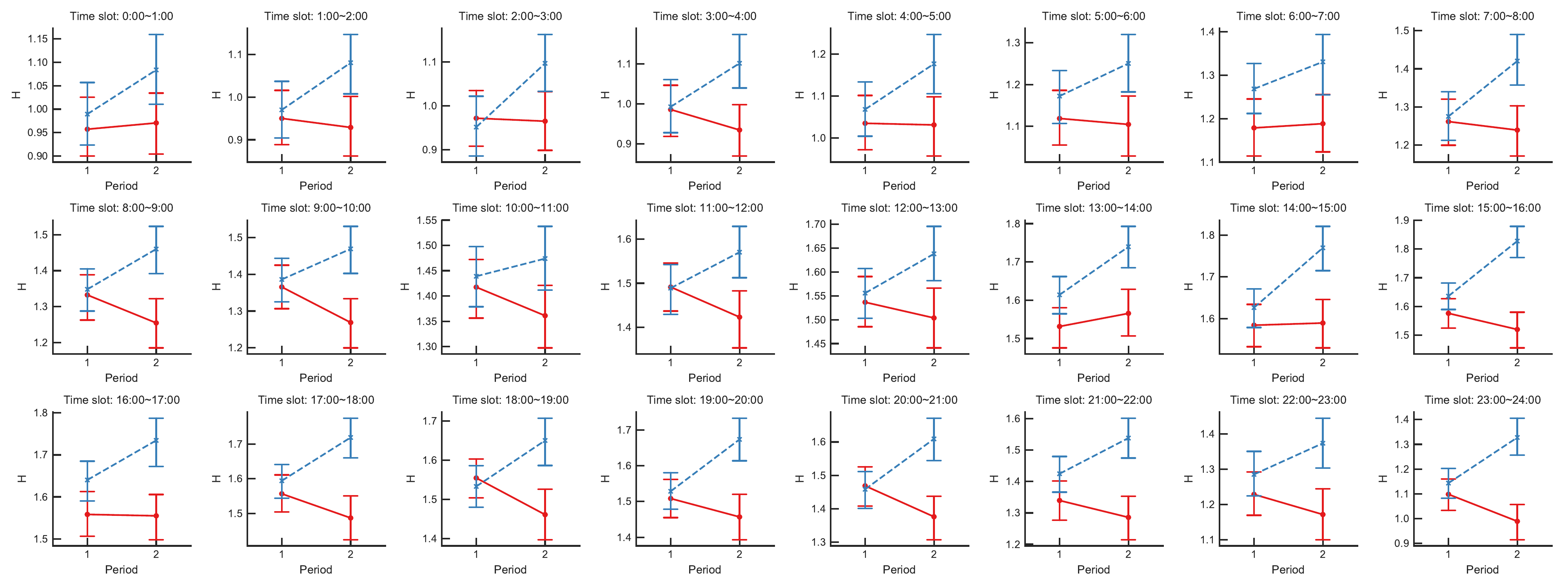}}\hfill{}

\caption{\label{fig:DailyEntropy-1-1-1}The mean cross-entropies of the distributions
of hourly time spending between players and churner community in \textit{Game
of Thrones Winter is Coming (GOT)} where the playing time is divided
by $m=2$ periods. The blue dashed line shows the non-churners, while
the red solid line shows the churners.}

\hfill{}\subfloat{\includegraphics[clip,width=0.8\paperwidth]{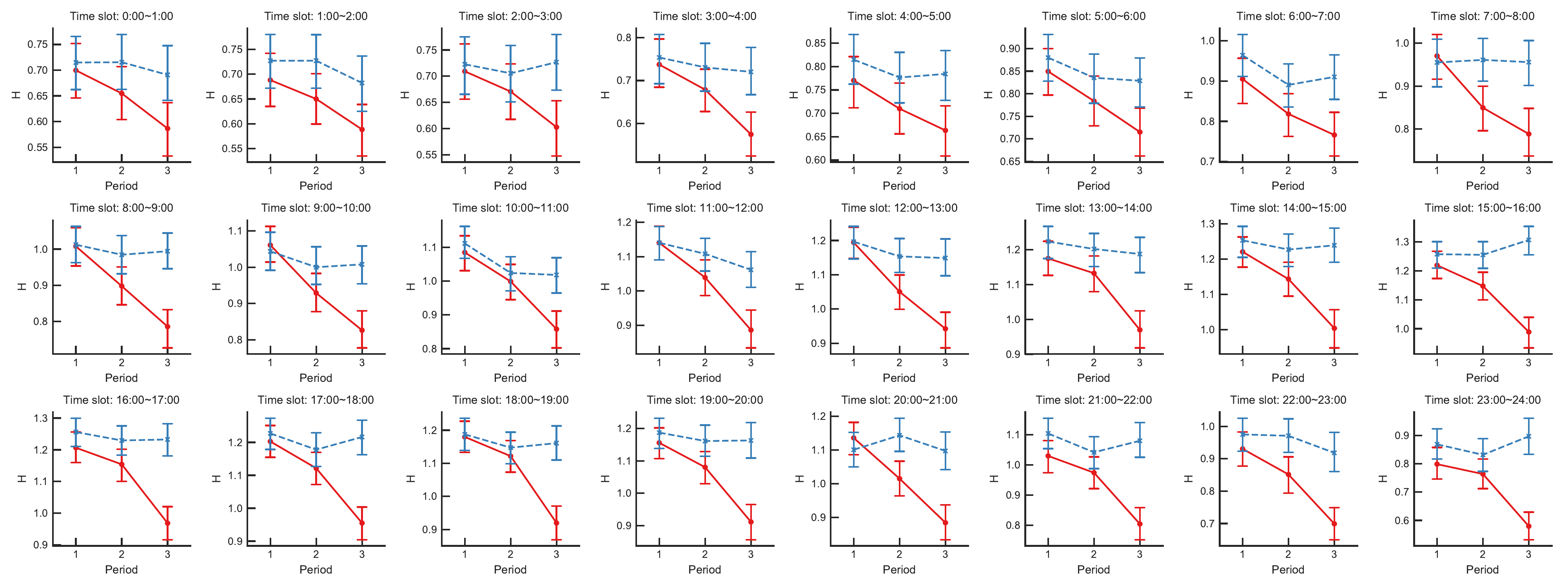}}\hfill{}

\caption{\label{fig:DailyCrossEntropyC-1-1-1}The mean cross-entropies of the
distributions of hourly time spending between players and churner
community in \textit{Game of Thrones Winter is Coming (GOT)} where
the playing time is divided by $m=3$ periods. The blue dashed line
shows the non-churners, while the red solid line shows the churners.}
\hfill{}\subfloat{\includegraphics[clip,width=0.8\paperwidth]{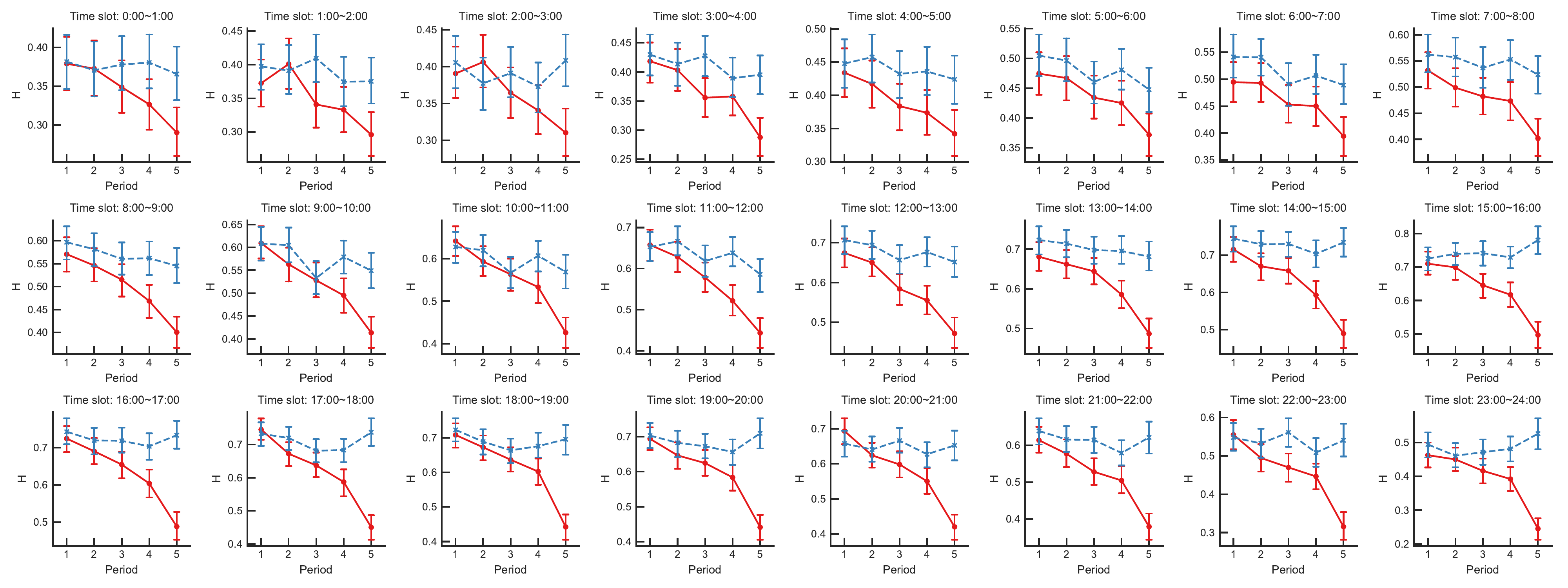}}\hfill{}

\caption{\label{fig:DailyCrossEntropyNC-1-1-1}The mean cross-entropies of
the distributions of hourly time spending between players and churner
community in \textit{Game of Thrones Winter is Coming (GOT)} where
the playing time is divided by $m=5$ periods. The blue dashed line
shows the non-churners, while the red solid line shows the churners.}
\end{figure*}

\begin{figure*}[t]
\hfill{}\subfloat{\centering{}\includegraphics[clip,width=0.8\paperwidth]{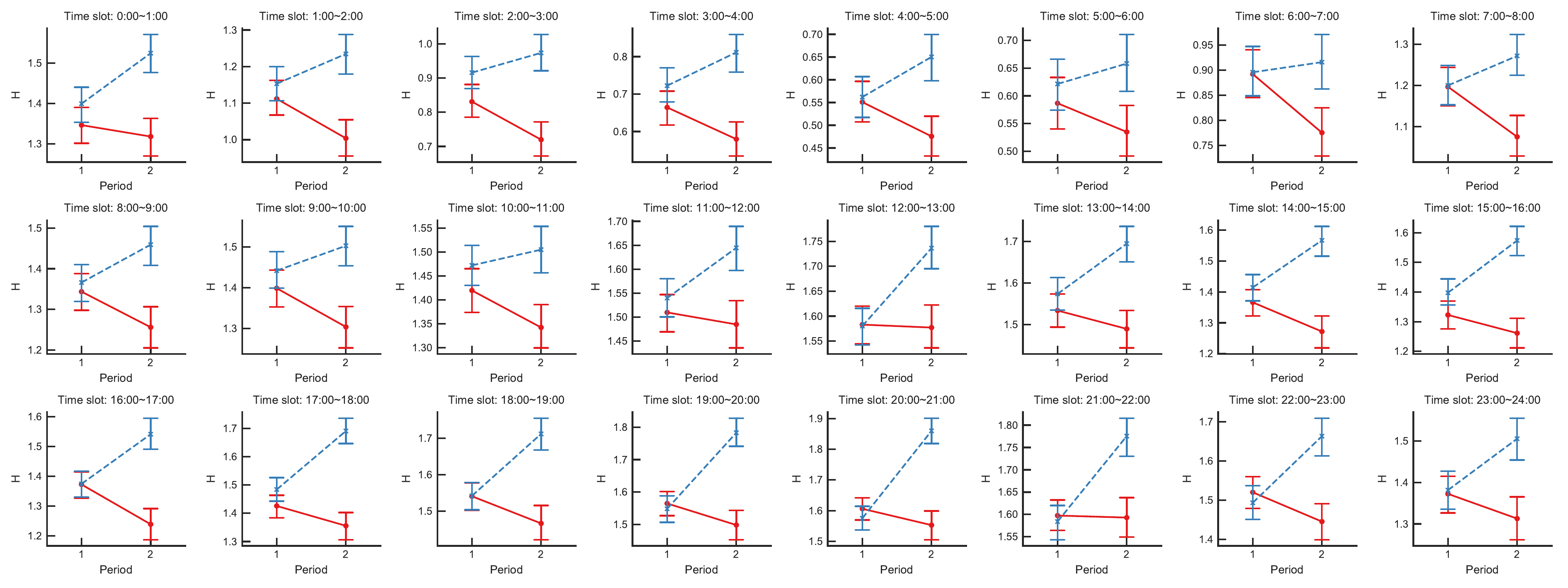}}\hfill{}

\caption{\label{fig:DailyEntropy-1-2-1}The mean cross-entropies of the distributions
of hourly time spending between players and churner community in \textit{Thirty-six
Stratagems} \textit{Mobile (TSM)} where the playing time is divided
by $m=2$ periods. The blue dashed line shows the non-churners, while
the red solid line shows the churners.}

\hfill{}\subfloat{\includegraphics[clip,width=0.8\paperwidth]{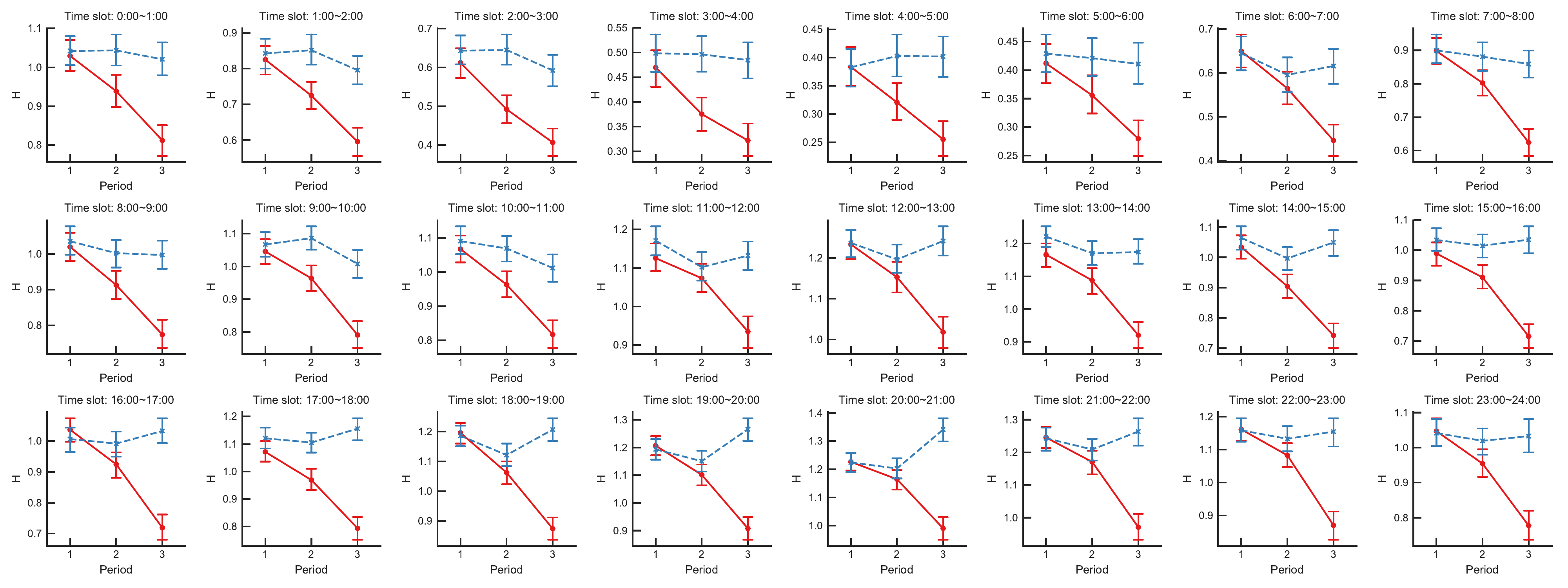}}\hfill{}

\caption{\label{fig:DailyCrossEntropyC-1-2-1}The mean cross-entropies of the
distributions of hourly time spending between players and churner
community in \textit{Thirty-six Stratagems} \textit{Mobile (TSM)}
where the playing time is divided by $m=3$ periods. The blue dashed
line shows the non-churners, while the red solid line shows the churners.}
\hfill{}\subfloat{\includegraphics[clip,width=0.8\paperwidth]{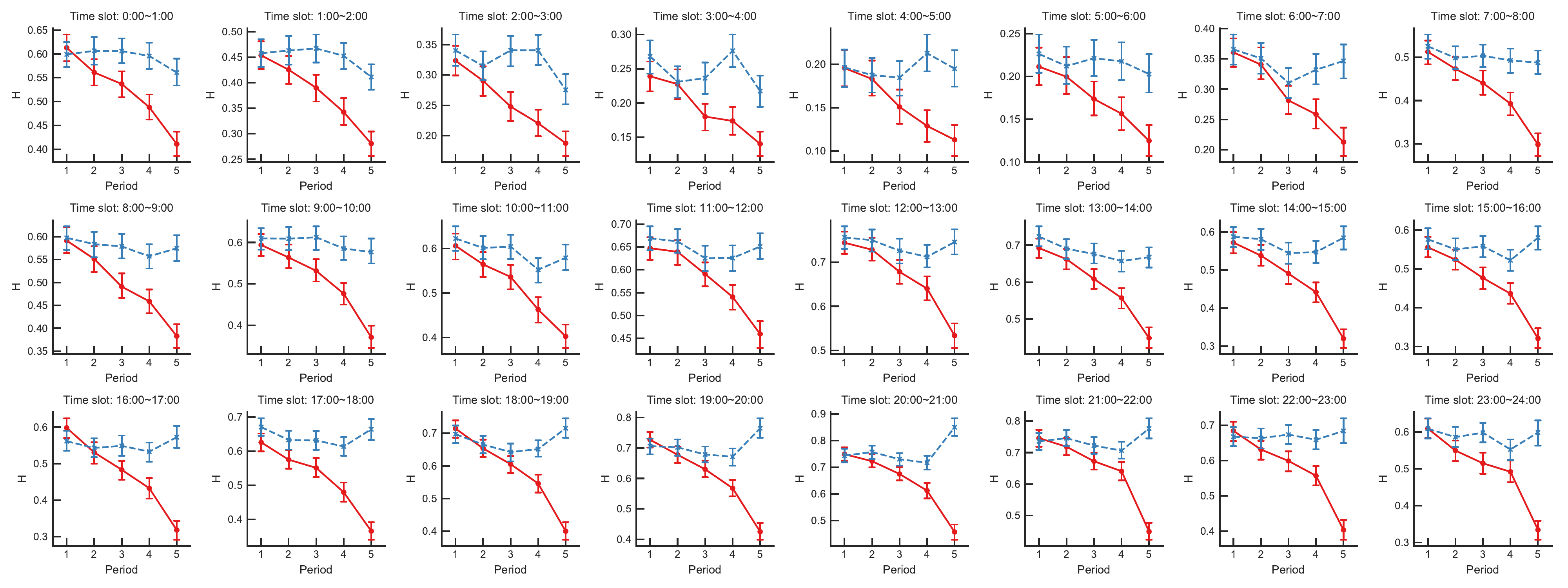}}\hfill{}

\caption{\label{fig:DailyCrossEntropyNC-1-2-1}The mean cross-entropies of
the distributions of hourly time spending between players and churner
community in \textit{Thirty-six Stratagems} \textit{Mobile (TSM) }where
the playing time is divided by $m=5$ periods. The blue dashed line
shows the non-churners, while the red solid line shows the churners.}
\end{figure*}

\begin{figure*}[t]
\hfill{}\subfloat{\centering{}\includegraphics[clip,width=0.8\paperwidth]{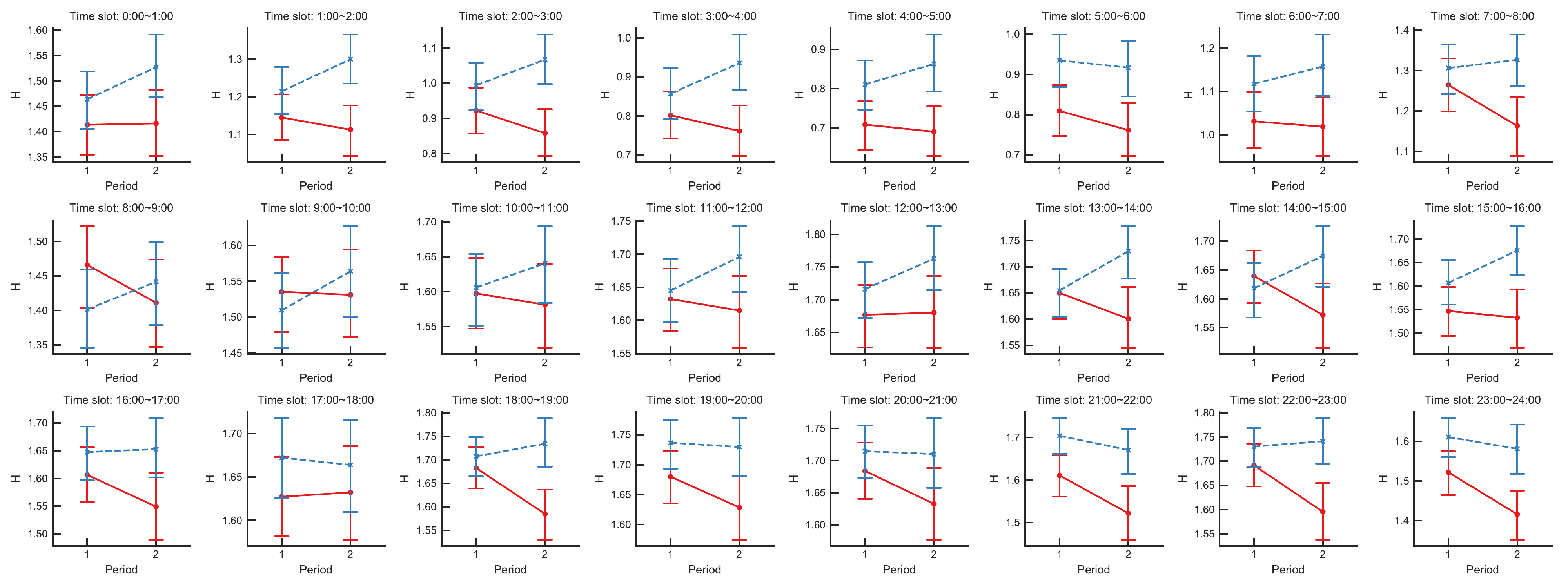}}\hfill{}

\caption{\label{fig:30}The mean cross-entropies of the distributions of hourly
time spending between players and churner community in \textit{Womanland
in Journey to the West (WJW) }where the playing time is divided by
$m=2$ periods. The blue dashed line shows the non-churners, while
the red solid line shows the churners.}

\hfill{}\subfloat{\includegraphics[clip,width=0.8\paperwidth]{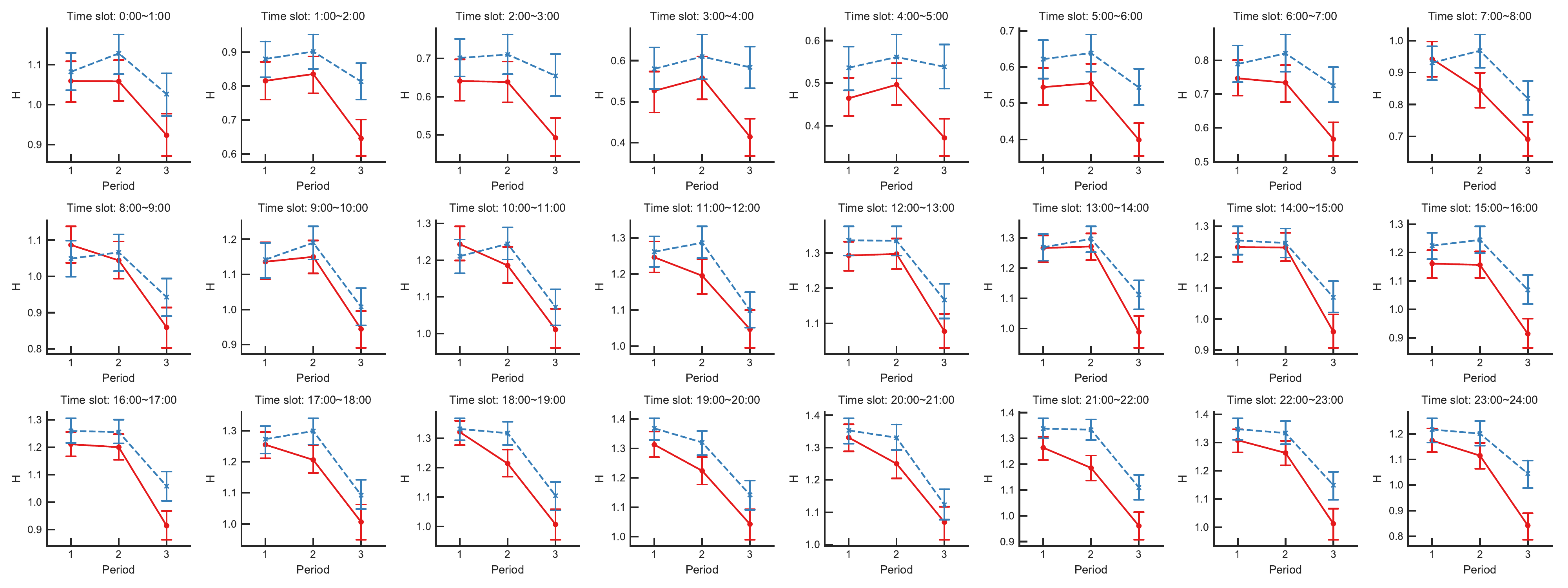}}\hfill{}

\caption{\label{fig:DailyCrossEntropyC-1-3-1}The mean cross-entropies of the
distributions of hourly time spending between players and churner
community in \textit{Womanland in Journey to the West (WJW) }where
the playing time is divided by $m=3$ periods. The blue dashed line
shows the non-churners, while the red solid line shows the churners.}
\hfill{}\subfloat{\includegraphics[clip,width=0.8\paperwidth]{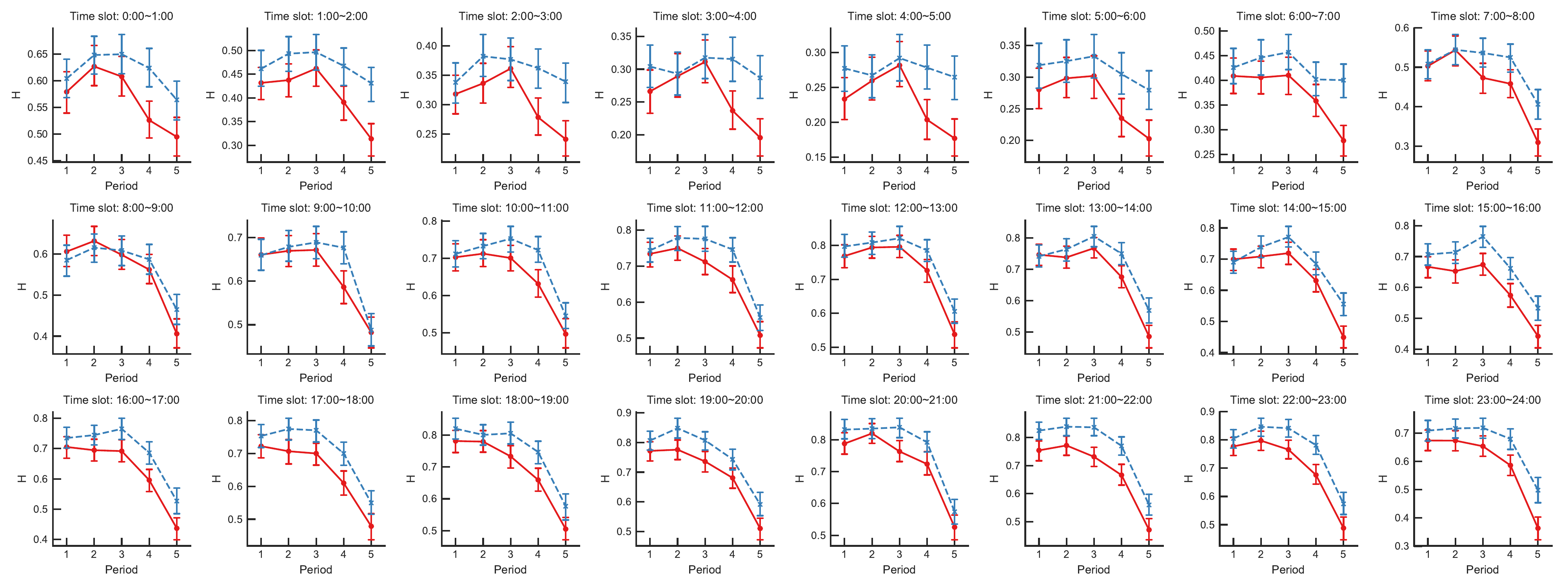}}\hfill{}

\caption{\label{fig:32}The mean cross-entropies of the distributions of hourly
time spending between players and churner community in \textit{Womanland
in Journey to the West (WJW) }where the playing time is divided by
$m=5$ periods. The blue dashed line shows the non-churners, while
the red solid line shows the churners.}
\end{figure*}

\begin{figure*}[t]
\hfill{}\subfloat{\centering{}\includegraphics[clip,width=0.8\paperwidth]{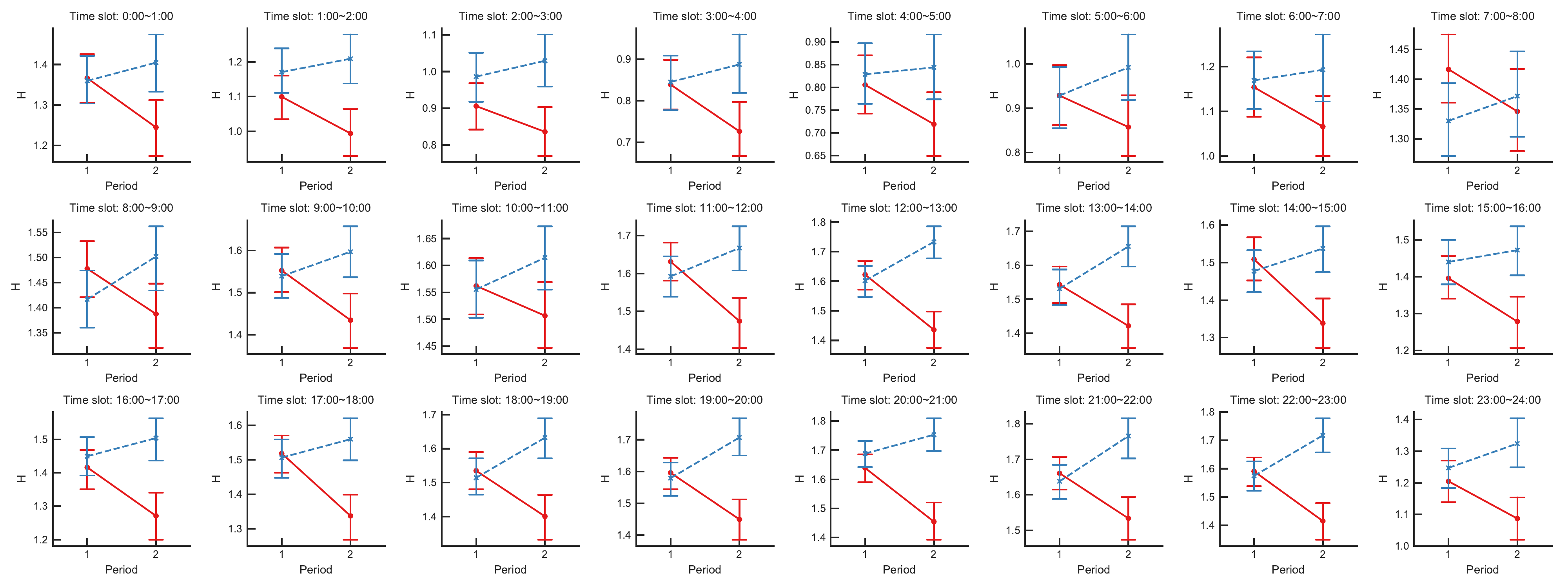}}\hfill{}

\caption{\label{fig:DailyEntropy-1-4-1}The mean cross-entropies of the distributions
of hourly time spending between players and churner community in \textit{Era
of Angels (EOA) }where the playing time is divided by $m=2$ periods.
The blue dashed line shows the non-churners, while the red solid line
shows the churners.}

\hfill{}\subfloat{\includegraphics[clip,width=0.8\paperwidth]{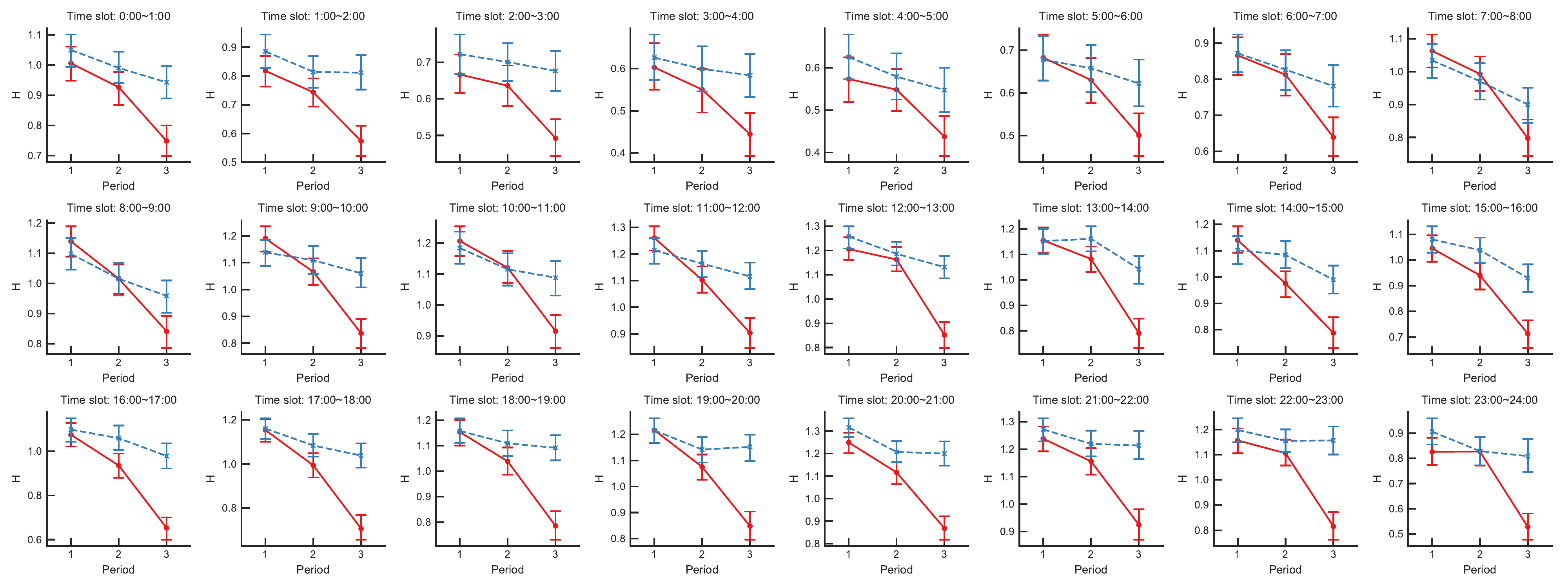}}\hfill{}

\caption{\label{fig:DailyCrossEntropyC-1-4-1}The mean cross-entropies of the
distributions of hourly time spending between players and churner
community in \textit{Era of Angels (EOA) }where the playing time is
divided by $m=3$ periods. The blue dashed line shows the non-churners,
while the red solid line shows the churners.}
\hfill{}\subfloat{\includegraphics[clip,width=0.8\paperwidth]{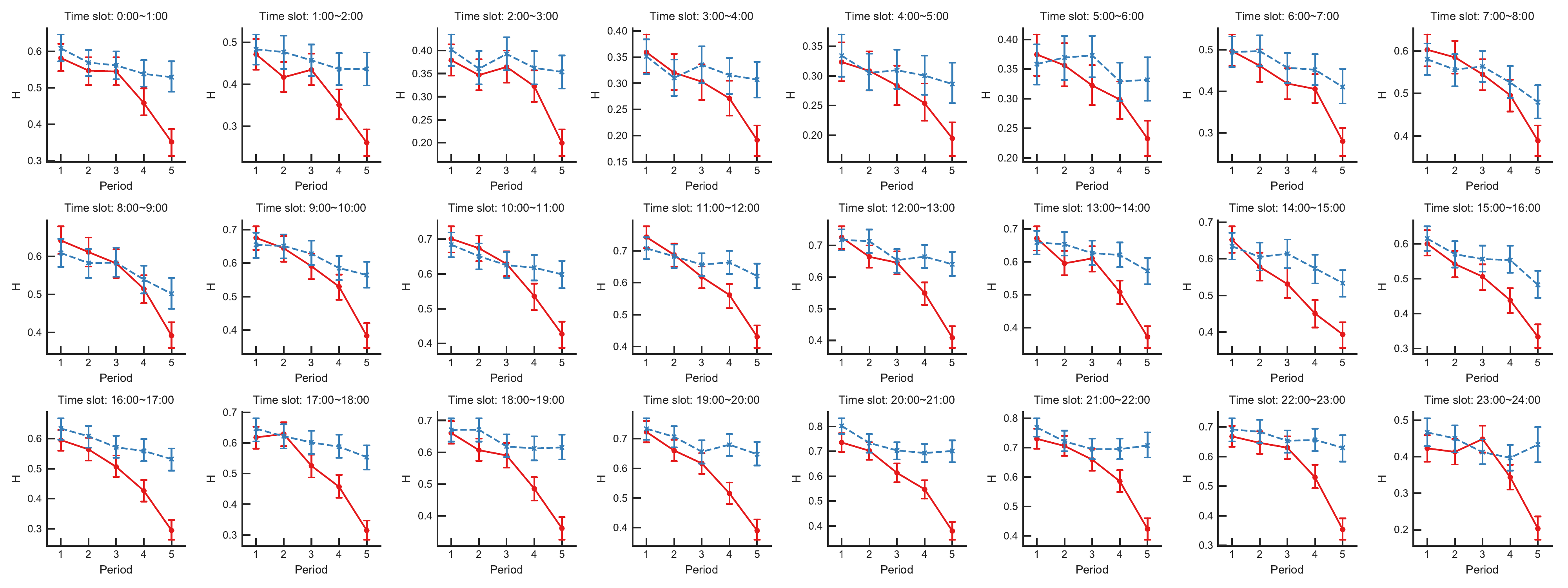}}\hfill{}

\caption{\label{fig:DailyCrossEntropyNC-1-4-1}The mean cross-entropies of
the distributions of hourly time spending between players and churner
community in \textit{Era of Angels (EOA) }where the playing time is
divided by $m=5$ periods. The blue dashed line shows the non-churners,
while the red solid line shows the churners.}
\end{figure*}

\begin{figure*}[t]
\hfill{}\subfloat{\centering{}\includegraphics[clip,width=0.8\paperwidth]{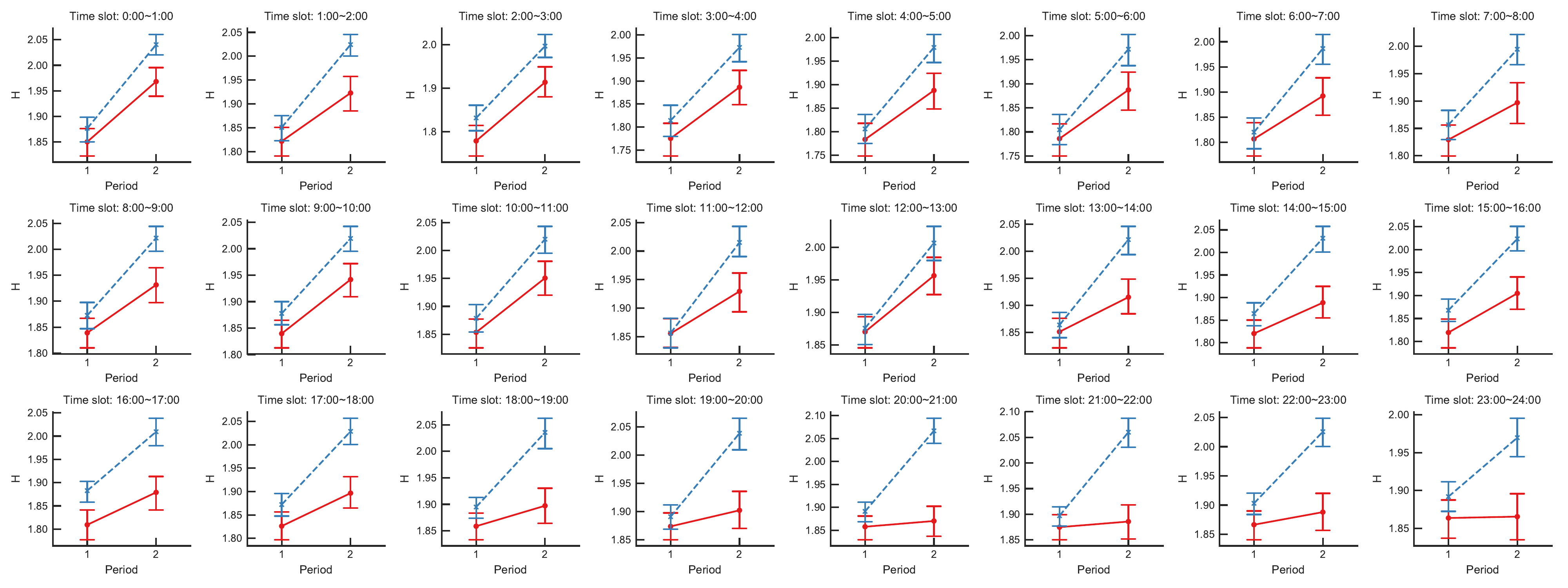}}\hfill{}

\caption{\label{fig:DailyEntropy-1-5-1}The mean cross-entropies of the distributions
of hourly time spending between players and churner community in \textit{League
of Angels II (LOA II) }where the playing time is divided by $m=2$
periods. The blue dashed line shows the non-churners, while the red
solid line shows the churners.}

\hfill{}\subfloat{\includegraphics[clip,width=0.8\paperwidth]{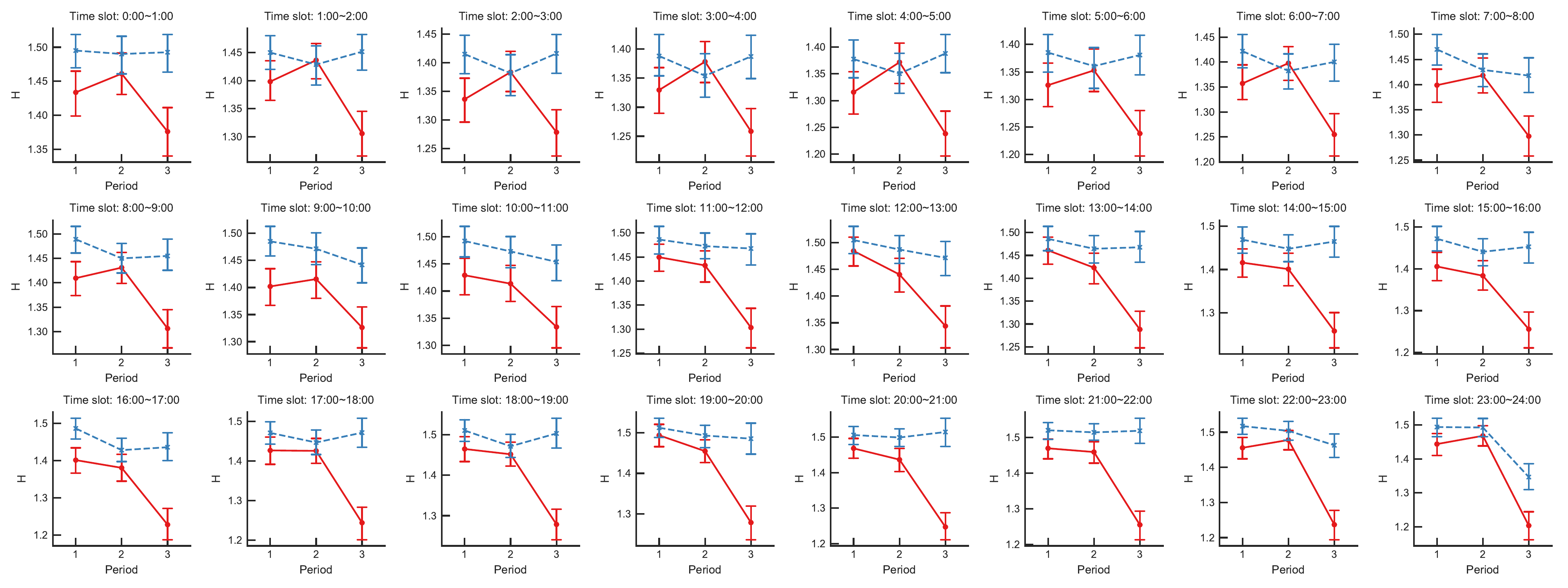}}\hfill{}

\caption{\label{fig:DailyCrossEntropyC-1-5-1}The mean cross-entropies of the
distributions of hourly time spending between players and churner
community in \textit{League of Angels II (LOA II) }where the playing
time is divided by $m=3$ periods. The blue dashed line shows the
non-churners, while the red solid line shows the churners.}
\hfill{}\subfloat{\includegraphics[clip,width=0.8\paperwidth]{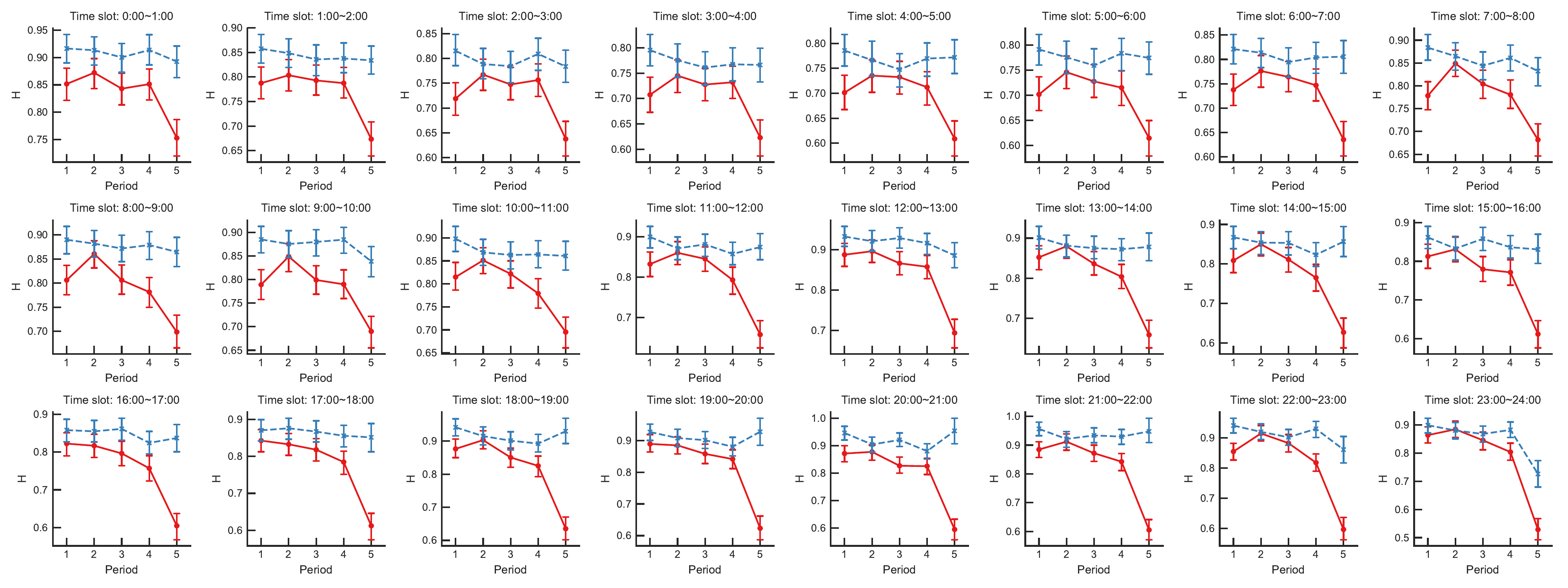}}\hfill{}

\caption{\label{fig:38}The mean cross-entropies of the distributions of hourly
time spending between players and churner community in \textit{League
of Angels II (LOA II) }where the playing time is divided by $m=5$
periods. The blue dashed line shows the non-churners, while the red
solid line shows the churners.}
\end{figure*}

For each game, we calculate the global playtime distributions and
cross-entropies using the training datasets. There are two kinds of
cross-entropies, churner-churners and nonchurner-churners. For example,
a nonchurner-churners cross-entropy is the cross-entropy between a
non-churner's $p_{\text{ind}}$ and the churner community's, e.g.,
$p_{\text{global}}^{\text{churner}}(d|k,r)$, for the case of hourly
feature. The mean cross-entropies and $95\%$ confidence intervals
are shown in Appendix \ref{sec:The-Entropy-Distributions} (Fig. \ref{fig:DailyCrossEntropy}
and Fig. \ref{fig:21} - Fig. \ref{fig:38}). We observe from Fig.
\ref{fig:DailyCrossEntropy} that there is no significant difference
between the mean cross-entropy of churner-churners and the mean cross-entropy
of nonchurner-churners when the daily playtime distributions are used,
for any number of periods $m$.

On the other hand, Fig. \ref{fig:21} - Fig. \ref{fig:38} show that
the mean hourly nonchurner-churners cross-entropies are higher than
those of churner-churners. It implies that the playtime pattern of
the non-churner is less similar to that of churners' community.

\begin{figure*}[t]
\hfill{}\subfloat{\centering{}\includegraphics[clip,width=0.8\paperwidth]{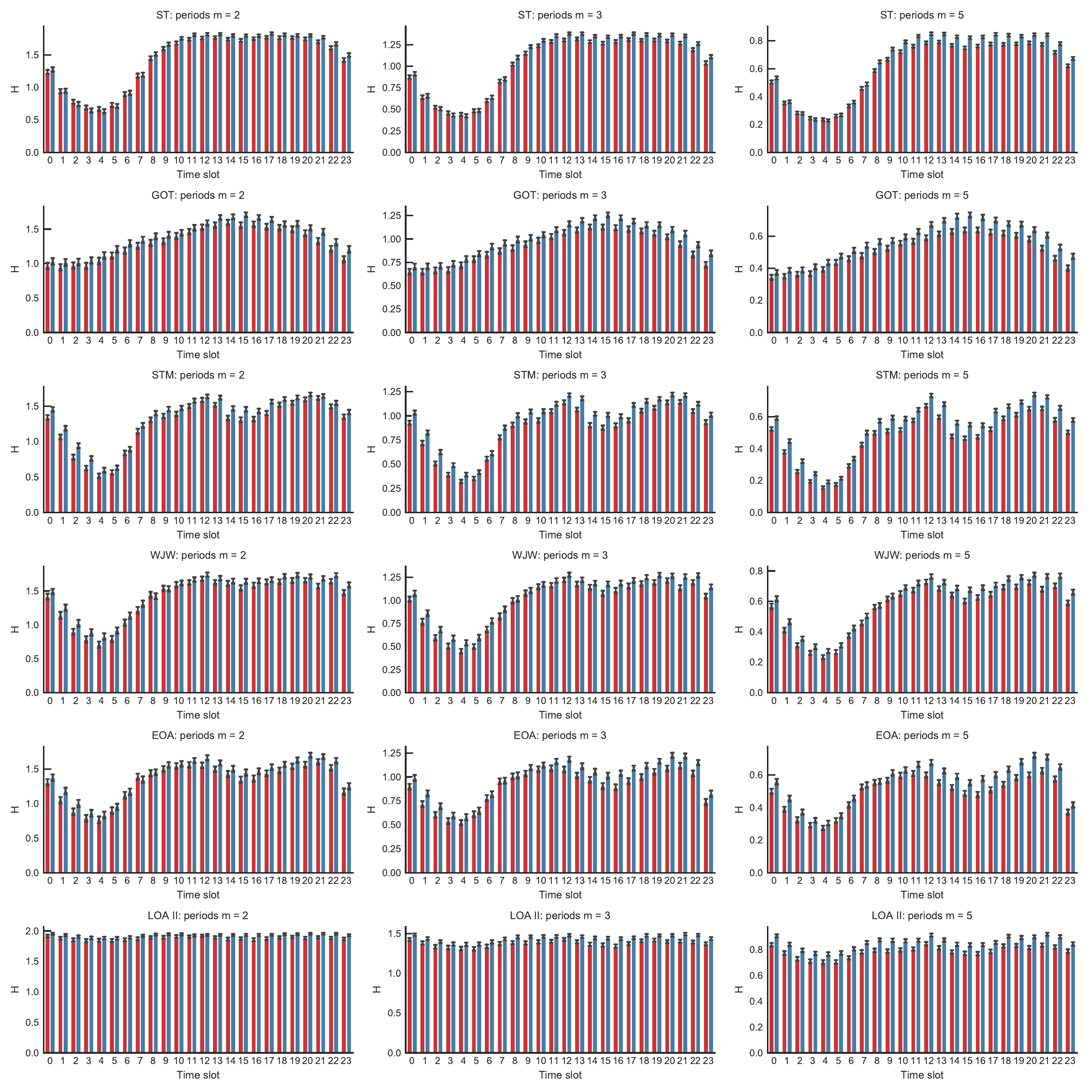}}\hfill{}

\caption{\label{fig:-1}The mean cross-entropies of the distributions of hourly
time spending between players and churner community in different time
slot (from 0:00 - 1:00 to 23:00 - 24:00, indexed from 0 - 23) in the
aforementioned games. The blue bar shows the non-churners, while the
red bar shows the churners.}
\end{figure*}

Further more, the difference is more significant in the hours from
early morning to late night and less significant from late night to
early morning, as seen in Fig. \ref{fig:-1}. These characteristics
of the hourly cross-entropy is similar to that of the hourly entropy
presented in the last subsection.

\subsection{Players' Playtime Regularity for Churn Prediction}

To summarize, churners and non-churners exhibit different playtime
regularities or patterns as captured by the entropies and cross-entropies
of the playtime distributions:
\begin{itemize}
\item Churners have lower entropies or larger playtime irregularity, as
well as a larger decrease in entropy than non-churners as time moves
on.
\item %
The playtime pattern of the non-churner is less similar to that of
churners' community. Non-churners have increasingly higher cross-entropies
(compared to the churner community) %
in hourly playtime distribution as time moves on. %
{} In other words, for cross-entropies of hourly playtime distributions,
nonchurner-churners is higher than churner-churners.
\end{itemize}
In the next two sections, we will utilize these differences to engineer
entropy-based features for churn prediction.

\section{Feature Engineering\label{sec:Feature-Engineering}}

The observation in Section \ref{sec:Mining-Player-Time} shows that
churners and non-churners exhibit different playtime regularity that
can be captured by the corresponding entropies. In this section, we
propose several features based on entropies that will be used for
churn prediction in the next section.

\subsection{Static Feature and Rate Feature}

For a given playtime distribution for player $u$ in $k$-th period
($k\leq m$), we define a function $f(u,k)$ representing the corresponding
entropy or cross-entropy. We call $f_{\text{}}(u,k)$ a static feature.
Based on the static feature, we define a rate feature: 
\begin{align*}
g_{\text{}}(u,k)= & \frac{f_{\text{}}(u,k)-f_{\text{}}(u,k+1)}{f_{\text{}}(u,k)}
\end{align*}
to capture the change in entropy/cross-entropy as time moves on where
$k<m$. Recall from the last section that churners exhibit smaller
entropies but with a greater entropy decrease as time increases. The
rate feature amplifies the differences between churners and non-churners.

\subsection{Feature Selection}

As seen in Section \prettyref{subsec:Period-Variation-Signals} and
Section \prettyref{subsec:Gaming-Community-Comparison}, churners
and non-churners exhibit differences in entropy of the daily playtime
distribution and of hourly playtime distribution, as well as in cross-entropy
of hourly playtime distribution. We therefore select four types of
features as follows:
\begin{itemize}
\item The combination of the static feature and rate feature of entropy
of the \emph{daily} playtime distributions. We call this type of features
the $1$st type of the proposed features.%
\item The combination of the static feature and rate feature of entropy
of the \emph{hourly} playtime distributions. We call this type of
features the 2nd type of the proposed features.
\item The combination of the static feature and rate feature of cross-entropy
of the \emph{hourly} playtime distributions. We call this type of
features the $3$rd type of the proposed features.
\item The combination of $1$st, $2$nd, and $3$rd types of features. We
call this type of features as the combined type of the proposed features.
\end{itemize}

\section{Churner Prediction\label{sec:Churner-Detection-(Building}}

In this section, we evaluate the efficacy of churn prediction using
entropy features with several typical classifiers.

\subsection{Evaluation Strategy}

\subsubsection{Baseline Features}

\begin{table*}[t]
\begin{centering}
\caption{\label{tab:Score-1}AUC for Different Proposed Features (Classifiers
Trained With the Data Where Playing Time is Divided by $m=2,3,5$
Periods)}
\par\end{centering}
\centering{}%
\begin{tabular}{c>{\raggedright}m{0.07\textwidth}ccccccccccccc}
\hline 
 &  &  & \multicolumn{4}{c}{Period $m=2$} & \multicolumn{4}{c}{Period $m=3$} & \multicolumn{4}{c}{Period $m=5$}\tabularnewline
\hline 
 &  &  & LR & SVM & DT & RF & LR & SVM & DT & RF & LR & SVM & DT & RF\tabularnewline
\hline 
\hline 
\multirow{7}{*}{\textit{ST}} & \multirow{4}{0.07\textwidth}{\centering{}Proposed Features} & 1st Type & 0.667 & 0.666 & 0.546 & 0.660 & 0.692 & 0.691 & 0.556 & 0.693 & 0.711 & 0.710 & 0.566 & 0.706\tabularnewline
\cline{3-15} \cline{4-15} \cline{5-15} \cline{6-15} \cline{7-15} \cline{8-15} \cline{9-15} \cline{10-15} \cline{11-15} \cline{12-15} \cline{13-15} \cline{14-15} \cline{15-15} 
 &  & 2nd Type & 0.648 & 0.648 & 0.556 & 0.668 & 0.667 & 0.668 & 0.576 & 0.676 & 0.689 & 0.687 & 0.602 & 0.681\tabularnewline
\cline{3-15} \cline{4-15} \cline{5-15} \cline{6-15} \cline{7-15} \cline{8-15} \cline{9-15} \cline{10-15} \cline{11-15} \cline{12-15} \cline{13-15} \cline{14-15} \cline{15-15} 
 &  & 3rd Type & 0.607 & 0.613 & 0.607 & 0.715 & 0.620 & 0.621 & 0.616 & 0.730 & 0.669 & 0.668 & 0.638 & 0.720\tabularnewline
\cline{3-15} \cline{4-15} \cline{5-15} \cline{6-15} \cline{7-15} \cline{8-15} \cline{9-15} \cline{10-15} \cline{11-15} \cline{12-15} \cline{13-15} \cline{14-15} \cline{15-15} 
 &  & Combined Type & 0.688 & 0.687 & 0.620 & 0.706 & 0.692 & 0.693 & 0.637 & 0.720 & 0.720 & 0.714 & 0.625 & \textbf{0.733}\tabularnewline
\cline{2-15} \cline{3-15} \cline{4-15} \cline{5-15} \cline{6-15} \cline{7-15} \cline{8-15} \cline{9-15} \cline{10-15} \cline{11-15} \cline{12-15} \cline{13-15} \cline{14-15} \cline{15-15} 
 & \multirow{3}{0.07\textwidth}{\centering{}Baseline Features} & Raw Data & 0.617 & 0.616 & 0.562 & 0.638 &  &  &  &  &  &  &  & \tabularnewline
\cline{3-15} \cline{4-15} \cline{5-15} \cline{6-15} \cline{7-15} \cline{8-15} \cline{9-15} \cline{10-15} \cline{11-15} \cline{12-15} \cline{13-15} \cline{14-15} \cline{15-15} 
 &  & 1st Type & 0.635 & 0.635 & 0.555 & 0.642 &  &  &  &  &  &  &  & \tabularnewline
\cline{3-15} \cline{4-15} \cline{5-15} \cline{6-15} \cline{7-15} \cline{8-15} \cline{9-15} \cline{10-15} \cline{11-15} \cline{12-15} \cline{13-15} \cline{14-15} \cline{15-15} 
 &  & 2nd Type & 0.585 & 0.584 & 0.522 & 0.569 &  &  &  &  &  &  &  & \tabularnewline
\hline 
\hline 
\multirow{7}{*}{\textit{GOT}} & \multirow{4}{0.07\textwidth}{\centering{}Proposed Features} & 1st Type & 0.682 & 0.682 & 0.545 & 0.672 & 0.720 & 0.717 & 0.576 & 0.702 & \textbf{0.741} & 0.740 & 0.586 & 0.720\tabularnewline
\cline{3-15} \cline{4-15} \cline{5-15} \cline{6-15} \cline{7-15} \cline{8-15} \cline{9-15} \cline{10-15} \cline{11-15} \cline{12-15} \cline{13-15} \cline{14-15} \cline{15-15} 
 &  & 2nd Type & 0.627 & 0.624 & 0.544 & 0.633 & 0.653 & 0.653 & 0.554 & 0.665 & 0.653 & 0.638 & 0.581 & 0.651\tabularnewline
\cline{3-15} \cline{4-15} \cline{5-15} \cline{6-15} \cline{7-15} \cline{8-15} \cline{9-15} \cline{10-15} \cline{11-15} \cline{12-15} \cline{13-15} \cline{14-15} \cline{15-15} 
 &  & 3rd Type & 0.631 & 0.636 & 0.600 & 0.708 & 0.643 & 0.644 & 0.601 & 0.725 & 0.677 & 0.668 & 0.584 & 0.714\tabularnewline
\cline{3-15} \cline{4-15} \cline{5-15} \cline{6-15} \cline{7-15} \cline{8-15} \cline{9-15} \cline{10-15} \cline{11-15} \cline{12-15} \cline{13-15} \cline{14-15} \cline{15-15} 
 &  & Combined Type & 0.643 & 0.650 & 0.595 & 0.714 & 0.666 & 0.654 & 0.589 & 0.731 & 0.674 & 0.670 & 0.604 & 0.729\tabularnewline
\cline{2-15} \cline{3-15} \cline{4-15} \cline{5-15} \cline{6-15} \cline{7-15} \cline{8-15} \cline{9-15} \cline{10-15} \cline{11-15} \cline{12-15} \cline{13-15} \cline{14-15} \cline{15-15} 
 & \multirow{3}{0.07\textwidth}{\centering{}Baseline Features} & Raw Data & 0.570 & 0.569 & 0.551 & 0.613 &  &  &  &  &  &  &  & \tabularnewline
\cline{3-15} \cline{4-15} \cline{5-15} \cline{6-15} \cline{7-15} \cline{8-15} \cline{9-15} \cline{10-15} \cline{11-15} \cline{12-15} \cline{13-15} \cline{14-15} \cline{15-15} 
 &  & 1st Type & 0.668 & 0.665 & 0.542 & 0.640 &  &  &  &  &  &  &  & \tabularnewline
\cline{3-15} \cline{4-15} \cline{5-15} \cline{6-15} \cline{7-15} \cline{8-15} \cline{9-15} \cline{10-15} \cline{11-15} \cline{12-15} \cline{13-15} \cline{14-15} \cline{15-15} 
 &  & 2nd Type & 0.594 & 0.593 & 0.510 & 0.556 &  &  &  &  &  &  &  & \tabularnewline
\hline 
\hline 
\multirow{7}{*}{\textit{STM}} & \multirow{4}{0.07\textwidth}{\centering{}Proposed Features} & 1st Type & 0.737 & 0.737 & 0.584 & 0.715 & 0.766 & 0.764 & 0.606 & 0.744 & \textbf{0.789} & 0.789 & 0.637 & 0.761\tabularnewline
\cline{3-15} \cline{4-15} \cline{5-15} \cline{6-15} \cline{7-15} \cline{8-15} \cline{9-15} \cline{10-15} \cline{11-15} \cline{12-15} \cline{13-15} \cline{14-15} \cline{15-15} 
 &  & 2nd Type & 0.701 & 0.699 & 0.577 & 0.675 & 0.735 & 0.734 & 0.582 & 0.659 & 0.737 & 0.739 & 0.603 & 0.684\tabularnewline
\cline{3-15} \cline{4-15} \cline{5-15} \cline{6-15} \cline{7-15} \cline{8-15} \cline{9-15} \cline{10-15} \cline{11-15} \cline{12-15} \cline{13-15} \cline{14-15} \cline{15-15} 
 &  & 3rd Type & 0.699 & 0.698 & 0.623 & 0.712 & 0.740 & 0.737 & 0.632 & 0.737 & 0.784 & 0.778 & 0.624 & 0.739\tabularnewline
\cline{3-15} \cline{4-15} \cline{5-15} \cline{6-15} \cline{7-15} \cline{8-15} \cline{9-15} \cline{10-15} \cline{11-15} \cline{12-15} \cline{13-15} \cline{14-15} \cline{15-15} 
 &  & Combined Type & 0.726 & 0.729 & 0.638 & 0.756 & 0.756 & 0.757 & 0.638 & 0.761 & 0.774 & 0.771 & 0.642 & 0.751\tabularnewline
\cline{2-15} \cline{3-15} \cline{4-15} \cline{5-15} \cline{6-15} \cline{7-15} \cline{8-15} \cline{9-15} \cline{10-15} \cline{11-15} \cline{12-15} \cline{13-15} \cline{14-15} \cline{15-15} 
 & \multirow{3}{0.07\textwidth}{\centering{}Baseline Features} & Raw Data & 0.672 & 0.677 & 0.559 & 0.669 &  &  &  &  &  &  &  & \tabularnewline
\cline{3-15} \cline{4-15} \cline{5-15} \cline{6-15} \cline{7-15} \cline{8-15} \cline{9-15} \cline{10-15} \cline{11-15} \cline{12-15} \cline{13-15} \cline{14-15} \cline{15-15} 
 &  & 1st Type & 0.715 & 0.717 & 0.594 & 0.686 &  &  &  &  &  &  &  & \tabularnewline
\cline{3-15} \cline{4-15} \cline{5-15} \cline{6-15} \cline{7-15} \cline{8-15} \cline{9-15} \cline{10-15} \cline{11-15} \cline{12-15} \cline{13-15} \cline{14-15} \cline{15-15} 
 &  & 2nd Type & 0.585 & 0.585 & 0.530 & 0.589 &  &  &  &  &  &  &  & \tabularnewline
\hline 
\hline 
\multirow{7}{*}{\textit{WJW}} & \multirow{4}{0.07\textwidth}{\centering{}Proposed Features} & 1st Type & 0.628 & 0.630 & 0.514 & 0.630 & 0.632 & 0.629 & 0.533 & 0.623 & \textbf{0.634} & 0.631 & 0.520 & 0.625\tabularnewline
\cline{3-15} \cline{4-15} \cline{5-15} \cline{6-15} \cline{7-15} \cline{8-15} \cline{9-15} \cline{10-15} \cline{11-15} \cline{12-15} \cline{13-15} \cline{14-15} \cline{15-15} 
 &  & 2nd Type & 0.621 & 0.629 & 0.521 & 0.627 & 0.585 & 0.588 & 0.508 & 0.618 & 0.585 & 0.584 & 0.552 & 0.603\tabularnewline
\cline{3-15} \cline{4-15} \cline{5-15} \cline{6-15} \cline{7-15} \cline{8-15} \cline{9-15} \cline{10-15} \cline{11-15} \cline{12-15} \cline{13-15} \cline{14-15} \cline{15-15} 
 &  & 3rd Type & 0.589 & 0.593 & 0.547 & 0.597 & 0.586 & 0.588 & 0.511 & 0.613 & 0.596 & 0.596 & 0.529 & 0.621\tabularnewline
\cline{3-15} \cline{4-15} \cline{5-15} \cline{6-15} \cline{7-15} \cline{8-15} \cline{9-15} \cline{10-15} \cline{11-15} \cline{12-15} \cline{13-15} \cline{14-15} \cline{15-15} 
 &  & Combined Type & 0.601 & 0.602 & 0.535 & 0.633 & 0.564 & 0.565 & 0.523 & 0.617 & 0.592 & 0.573 & 0.531 & 0.612\tabularnewline
\cline{2-15} \cline{3-15} \cline{4-15} \cline{5-15} \cline{6-15} \cline{7-15} \cline{8-15} \cline{9-15} \cline{10-15} \cline{11-15} \cline{12-15} \cline{13-15} \cline{14-15} \cline{15-15} 
 & \multirow{3}{0.07\textwidth}{\centering{}Baseline Features} & Raw Data & 0.557 & 0.529 & 0.505 & 0.605 &  &  &  &  &  &  &  & \tabularnewline
\cline{3-15} \cline{4-15} \cline{5-15} \cline{6-15} \cline{7-15} \cline{8-15} \cline{9-15} \cline{10-15} \cline{11-15} \cline{12-15} \cline{13-15} \cline{14-15} \cline{15-15} 
 &  & 1st Type & 0.618 & 0.626 & 0.524 & 0.620 &  &  &  &  &  &  &  & \tabularnewline
\cline{3-15} \cline{4-15} \cline{5-15} \cline{6-15} \cline{7-15} \cline{8-15} \cline{9-15} \cline{10-15} \cline{11-15} \cline{12-15} \cline{13-15} \cline{14-15} \cline{15-15} 
 &  & 2nd Type & 0.597 & 0.596 & 0.510 & 0.596 &  &  &  &  &  &  &  & \tabularnewline
\hline 
\hline 
\multirow{7}{*}{\textit{EOA}} & \multirow{4}{0.07\textwidth}{\centering{}Proposed Features} & 1st Type & 0.692 & 0.686 & 0.585 & 0.672 & 0.716 & 0.708 & 0.575 & 0.694 & \textbf{0.725} & 0.717 & 0.616 & 0.713\tabularnewline
\cline{3-15} \cline{4-15} \cline{5-15} \cline{6-15} \cline{7-15} \cline{8-15} \cline{9-15} \cline{10-15} \cline{11-15} \cline{12-15} \cline{13-15} \cline{14-15} \cline{15-15} 
 &  & 2nd Type & 0.651 & 0.646 & 0.547 & 0.633 & 0.658 & 0.667 & 0.564 & 0.651 & 0.632 & 0.630 & 0.595 & 0.645\tabularnewline
\cline{3-15} \cline{4-15} \cline{5-15} \cline{6-15} \cline{7-15} \cline{8-15} \cline{9-15} \cline{10-15} \cline{11-15} \cline{12-15} \cline{13-15} \cline{14-15} \cline{15-15} 
 &  & 3rd Type & 0.654 & 0.652 & 0.570 & 0.669 & 0.667 & 0.667 & 0.590 & 0.670 & 0.677 & 0.671 & 0.613 & 0.663\tabularnewline
\cline{3-15} \cline{4-15} \cline{5-15} \cline{6-15} \cline{7-15} \cline{8-15} \cline{9-15} \cline{10-15} \cline{11-15} \cline{12-15} \cline{13-15} \cline{14-15} \cline{15-15} 
 &  & Combined Type & 0.678 & 0.678 & 0.585 & 0.683 & 0.670 & 0.675 & 0.594 & 0.695 & 0.650 & 0.644 & 0.620 & 0.684\tabularnewline
\cline{2-15} \cline{3-15} \cline{4-15} \cline{5-15} \cline{6-15} \cline{7-15} \cline{8-15} \cline{9-15} \cline{10-15} \cline{11-15} \cline{12-15} \cline{13-15} \cline{14-15} \cline{15-15} 
 & \multirow{3}{0.07\textwidth}{\centering{}Baseline Features} & Raw Data & 0.611 & 0.616 & 0.540 & 0.641 &  &  &  &  &  &  &  & \tabularnewline
\cline{3-15} \cline{4-15} \cline{5-15} \cline{6-15} \cline{7-15} \cline{8-15} \cline{9-15} \cline{10-15} \cline{11-15} \cline{12-15} \cline{13-15} \cline{14-15} \cline{15-15} 
 &  & 1st Type & 0.640 & 0.628 & 0.603 & 0.676 &  &  &  &  &  &  &  & \tabularnewline
\cline{3-15} \cline{4-15} \cline{5-15} \cline{6-15} \cline{7-15} \cline{8-15} \cline{9-15} \cline{10-15} \cline{11-15} \cline{12-15} \cline{13-15} \cline{14-15} \cline{15-15} 
 &  & 2nd Type & 0.538 & 0.539 & 0.544 & 0.606 &  &  &  &  &  &  &  & \tabularnewline
\hline 
\hline 
\multirow{7}{*}{\textit{LOA II}} & \multirow{4}{0.07\textwidth}{\centering{}Proposed Features} & 1st Type & 0.679 & 0.677 & 0.559 & 0.680 & 0.698 & 0.697 & 0.578 & 0.682 & 0.703 & 0.704 & 0.579 & 0.700\tabularnewline
\cline{3-15} \cline{4-15} \cline{5-15} \cline{6-15} \cline{7-15} \cline{8-15} \cline{9-15} \cline{10-15} \cline{11-15} \cline{12-15} \cline{13-15} \cline{14-15} \cline{15-15} 
 &  & 2nd Type & 0.649 & 0.649 & 0.559 & 0.651 & 0.630 & 0.645 & 0.552 & 0.663 & 0.642 & 0.650 & 0.577 & 0.665\tabularnewline
\cline{3-15} \cline{4-15} \cline{5-15} \cline{6-15} \cline{7-15} \cline{8-15} \cline{9-15} \cline{10-15} \cline{11-15} \cline{12-15} \cline{13-15} \cline{14-15} \cline{15-15} 
 &  & 3rd Type & 0.618 & 0.634 & 0.586 & 0.711 & 0.628 & 0.624 & 0.579 & 0.731 & 0.661 & 0.661 & 0.588 & 0.727\tabularnewline
\cline{3-15} \cline{4-15} \cline{5-15} \cline{6-15} \cline{7-15} \cline{8-15} \cline{9-15} \cline{10-15} \cline{11-15} \cline{12-15} \cline{13-15} \cline{14-15} \cline{15-15} 
 &  & Combined Type & 0.645 & 0.649 & 0.591 & 0.736 & 0.633 & 0.631 & 0.594 & 0.735 & 0.646 & 0.645 & 0.612 & \textbf{0.739}\tabularnewline
\cline{2-15} \cline{3-15} \cline{4-15} \cline{5-15} \cline{6-15} \cline{7-15} \cline{8-15} \cline{9-15} \cline{10-15} \cline{11-15} \cline{12-15} \cline{13-15} \cline{14-15} \cline{15-15} 
 & \multirow{3}{0.07\textwidth}{\centering{}Baseline Features} & Raw Data & 0.548 & 0.544 & 0.504 & 0.636 &  &  &  &  &  &  &  & \tabularnewline
\cline{3-15} \cline{4-15} \cline{5-15} \cline{6-15} \cline{7-15} \cline{8-15} \cline{9-15} \cline{10-15} \cline{11-15} \cline{12-15} \cline{13-15} \cline{14-15} \cline{15-15} 
 &  & 1st Type & 0.600 & 0.597 & 0.526 & 0.639 &  &  &  &  &  &  &  & \tabularnewline
\cline{3-15} \cline{4-15} \cline{5-15} \cline{6-15} \cline{7-15} \cline{8-15} \cline{9-15} \cline{10-15} \cline{11-15} \cline{12-15} \cline{13-15} \cline{14-15} \cline{15-15} 
 &  & 2nd Type & 0.616 & 0.615 & 0.517 & 0.603 &  &  &  &  &  &  &  & \tabularnewline
\hline 
\end{tabular}
\end{table*}
To evaluate the effectiveness of our proposed features, we use following
baseline features for comparison:
\begin{itemize}
\item The raw data of the playtime distribution of players.
\item The daily total time spent, which we call the $1$st type of baseline
features.
\item The combined feature of total time spent, the last day of login, and
number of time slots played. This baseline feature is designed based
on Recency, Frequency, Monetary Value %
(RFM) model\cite{Lee_2005JoMR_RFMCLVUsingIsoValueCurvesCustomerBaseAnalysis}.
We call this type of feature as the $2$nd type of baseline features.
\end{itemize}

\subsubsection{AUC Evaluation}

Given the binary nature of prediction/classification, we use the area
under the Receiver-Operating-Characteristics (ROC) \cite{Fawcett_2006PRL_IntroductionROCAnalysis}
curve (AUC) \cite{Bradley_1997PR_UseAreaROCCurveEvaluationMachineLearningAlgorithms}
to evaluate the overall performance of a classifier. Note that, AUC
has been used as a metric to evaluate the performance of a certain
churn prediction model in many previous work, such as \cite{Faltings_20142ICCIG_ChurnPredictionHighvaluePlayersCasualSocialGames,Tsuzuki_2015ITCIAG_ChurnPredictionOnlineGamesUsingPlayersLoginRecordsFrequencyAnalysisApproach,Cowling_20152ICCIGC_PredictingPlayerDisengagementFirstPurchaseEventfrequencyBasedDataRepresentation,Hitchens_20162ICCIG_PredictingPlayerChurnDestinyHiddenMarkovModelsApproachPredictingPlayerDepartureMajorOnlineGame}.

\subsection{Evaluation Results}

We use multiple classifiers (Logistic Regression (LR), support vector
machine (SVM), Random Forests (RF)) and the aforementioned features
for churn prediction. For each game, the data set is equally divided
into a training data set and a test data set as we mention in Section
\ref{sec:Mining-Player-Time}%
. We train classifiers using the training data set and evaluate their
performance using the test data set. We evaluate the performance of
the classifiers that are trained with the data of the entire $m$
playing periods on the test data set. The results are shown in Table.\ref{tab:Score-1}
(Not that: the selected baseline features are not related to the choice
of period $m$).%
{} The best AUC score for each game is shown in bold font in Table.\ref{tab:Score-1}.

In our previous work \cite{Liu_20192ICGC_MiningPlayerIngameTimeSpendingRegularityChurnPredictionFreeOnlineGames},
the proposed entropy features have the best performance as opposed
to\textit{ }the baseline features for \textit{Thirty-six Stratagems}
and \textit{Thirty-six Stratagems Mobile}. In this paper, we see that,
our proposed entropy features have the best performance with the highest
AUC for all six games, as opposed to\textit{ }the baseline features.
We also find that the highest AUC is achieved when number of periods
is $m=5$ the for all these games, except \textit{League of Angels
II (LOA II)} where the highest AUC is achieved when number of periods
is $m=3$. This implies that a finer division on playtime may lead
to a better predictive power. It's worth noticing that for \textit{Womanland
in Journey to the West (WJW),} the achieved highest AUC is the lowest
among all games (still higher than the AUC of baseline features).
This is consistent with our observation in Fig. \ref{fig:DailyEntropy},
Fig. \ref{fig:12} - Fig. \ref{fig:14} and Fig. \ref{fig:30} -
Fig. \ref{fig:32}: The entropy/cross-entropy difference between churner
and non-churner is less significant here than in other games.

The evaluation shows that our proposed entropy features outperform
the baseline features in churn prediction, and indeed capture the
differences in the playtime between churners and non-churners. In
other words, although players' playtime serves as an important feature
in previous work, our proposed entropy features could utilize the
information extracted from players' playtime more effectively.

\section{Conclusions\label{sec:Conclusions}}

In this paper, we address the problem of better exploiting the information
extracted from players' playtime records for predicting churners in
free online games. The goal is to characterize and mine players' playtime
regularity among churners and non-churners. We consider long-term
players and understand their playtime regularity among churners and
non-churners. We observe that there is a significant difference between
churners and non-churners in terms of entropies exhibited in playtime
regularity. Based on this observation, we propose some prediction
models using new features extracted from mining players' playtime
regularity. After experiments are conducted, the corresponding result
shows that our proposed entropy/cross-entropy features are better
at predicting churners, compared to the baseline features. This implies
our proposed features could exploit the information extracted from
players' playtime records more effectively. It sheds light on a better
way of utilizing player's playtime records. Thus, game companies can
benefit from our algorithms by exploiting the information extracted
from players' playtime records more effectively from the database.
Our findings also help game developers design better in-game mechanisms
by increasing players' playtime regularity to reduce churn rate.

\section*{Acknowledgment}

We thank Scott Holman from the CU Boulder Writing Center for his feedback
and support during the writing process.

\bibliographystyle{IEEEtran}
\bibliography{IEEEabrv,IEEEexample,Data_Analytics_for_Players}

\appendices{}

\section{The Entropy/Cross-entropy Distribution on a Granularity of Each Day/Hour
for Player Time Spending in Each Aforementioned Game\label{sec:The-Entropy-Distributions}}

\begin{figure*}[tbh]
\hfill{}\subfloat{\centering{}\includegraphics[clip,width=0.7\paperwidth]{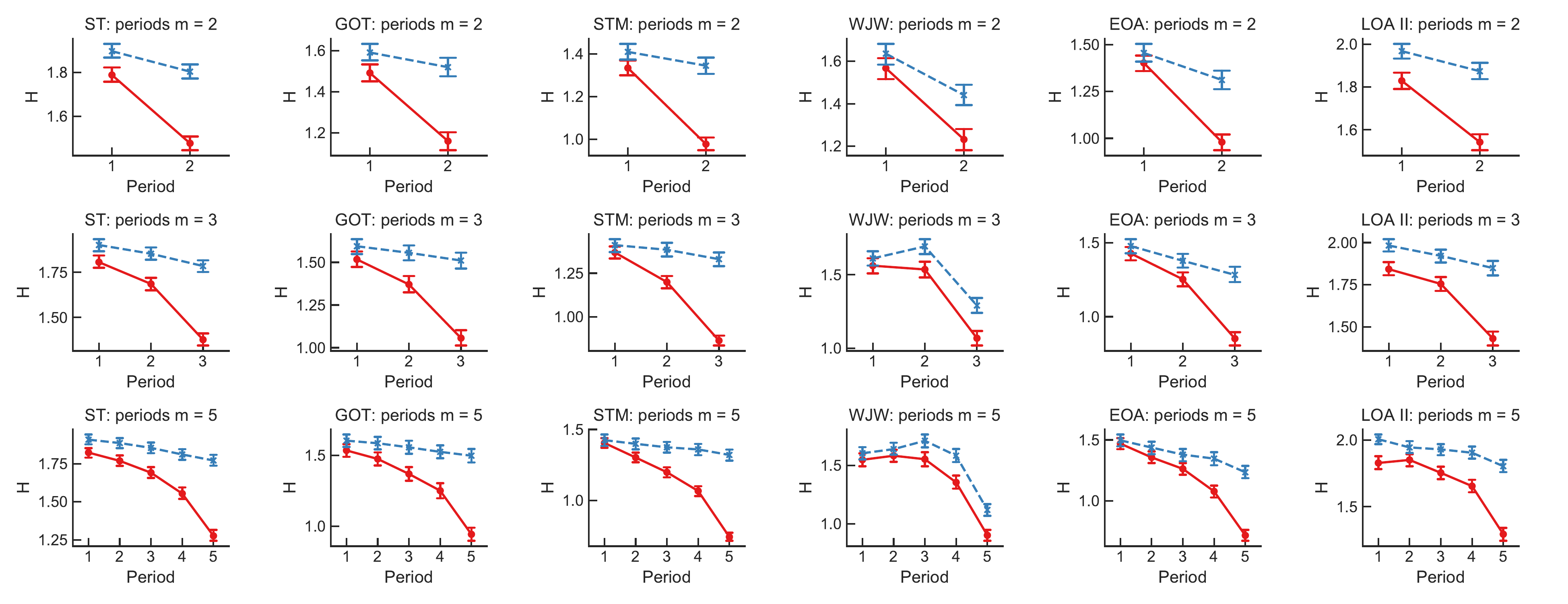}}\hfill{}

\caption{\label{fig:DailyEntropy}The mean entropies of the distributions of
daily time spending of aforementioned games players in different periods.
The blue dashed line shows the non-churners, while the red solid line
shows the churners.}
\hfill{}\subfloat{\centering{}\includegraphics[clip,width=0.7\paperwidth]{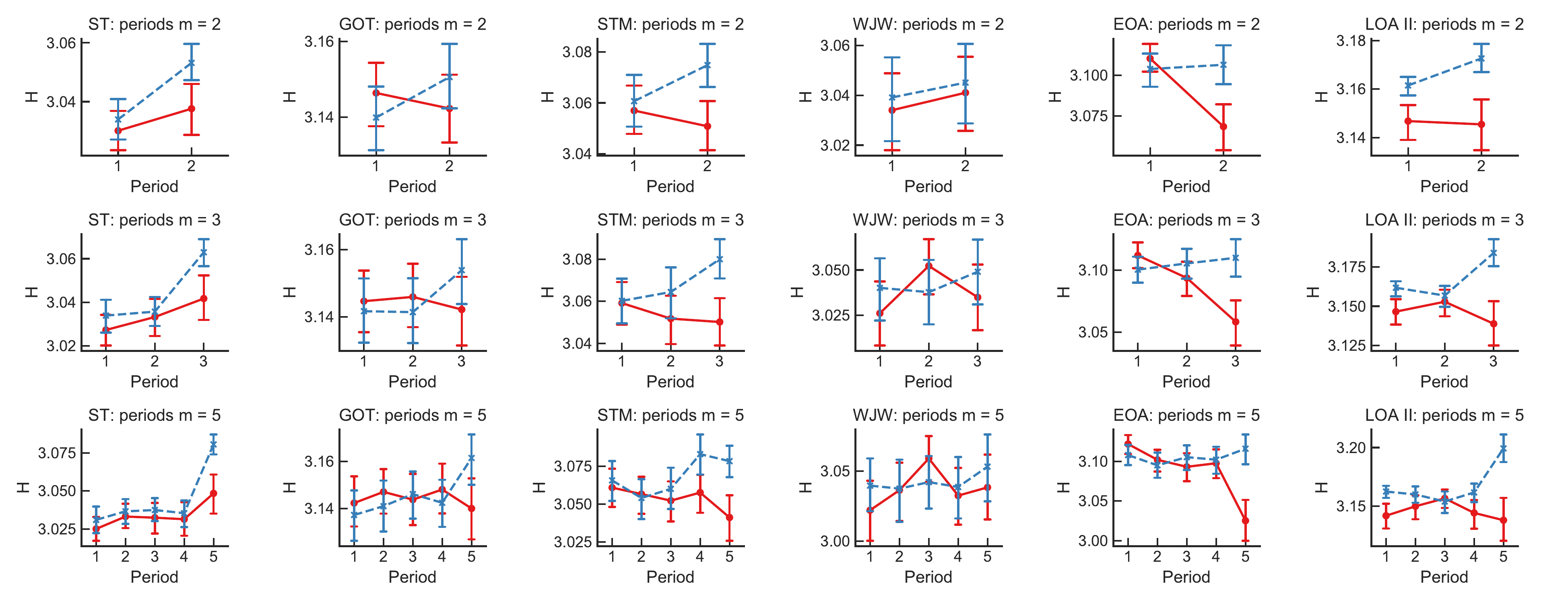}}\hfill{}

\caption{\label{fig:DailyCrossEntropy}The mean cross-entropies of the distributions
of daily time spending between players and churner community in aforementioned
games. The blue dashed line shows the non-churners, while the red
solid line shows the churners.}
\end{figure*}

\end{document}